\documentclass{aa}
\usepackage[varg]{txfonts}
\usepackage[utf8]{inputenc}
\usepackage{graphicx}
\usepackage{array}
\usepackage{hyperref}
\usepackage{multirow}
\usepackage{natbib}
\usepackage{enumitem}
\usepackage{pifont}
\usepackage{xfrac} 
\usepackage{nicefrac}
\usepackage{mathtools}
\usepackage{ulem}

\hypersetup{
    colorlinks=true,
    linkcolor=blue,
    citecolor=blue
}
\usepackage{color}

\begin{document}

\title{A graph-theory-based multi-scale analysis of hierarchical cascade in molecular clouds}
\titlerunning{A graph theory-based multi-scale analysis of hierarchical cascade}
\subtitle{Application to the NGC$~$2264 region}
\author{Thomasson B.$^{1}$
\and Joncour I.$^{1}$
\and Moraux E.$^{1}$
\and Crespelle C.$^{2}$
\and Motte F.$^{1}$
\and Pouteau Y.$^{1}$
\and Nony T.$^{3}$}

\institute{IPAG - Université Grenoble-Alpes
\and LIP - Université Claude Bernard Lyon 1
\and Instituto de Radioastronomía y Astrofísica, Universidad Nacional Autónoma de México, Apdo. Postal 3-72, 58089 Morelia, Michoacán, Mexico}

\date{Received <date> /
Accepted <date>}

\abstract {The spatial properties of small star clusters suggest that they may originate from a fragmentation cascade starting from molecular cloud, of which there might be traces found at spatial scales up to a few tens of thousands of astronomical units (kAU).} {Our goal is to investigate the multi-scale spatial structure of gas clumps, to probe the existence of a hierarchical cascade over a range of characteristic spatial scales, and to evaluate its possible link with star production in terms of multiplicity.} {From the Herschel emission maps of NGC$~$2264 at [70, 160, 250, 350, 500]$~\mu$m, clumps are extracted using \texttt{getsf} software at each of the associated spatial resolutions (respectively [8.4, 13.5, 18.2, 24.9, 36.3]"). Using the spatial distribution of these clumps and the class 0/I young stellar object (YSO) from Spitzer data, we developed a graph-theoretic analysis to represent the multi-scale structure of the cloud as a connected network. This network is organised in levels, and each level represents a characteristic scale among the available spatial scales. A link is created between two nodes which could be either a clump or a YSO from two different levels if their footprints overlap with each other. A parent node is then associated with a child node from a lower scale. The way in which the network subdivides scale after scale is compared with a geometric model that we have developed. This model generates extended objects that have a particularity in that they are geometrically constrained and subdivide along the scales following a fractal law. This graph-theoretic representation allows us to develop new statistical metrics and tools aiming at characterising, in a quantitative way, the multi-scale nature of molecular clouds.}{We obtain three classes of multi-scale structure in NGC$~$2264 according to the number of nodes produced at the deepest level (called graph-sinks): hierarchical (several graph-sinks), linear (a single graph-sink with at most a single parent at each level), and isolated (no connection to any other node). The class of structure is strongly correlated with the column density $N_{\rm H_2}$ of NGC$~$2264. The hierarchical structures dominate the regions whose column density exceeds N$_{\rm H_2} = 6 \times 10^{22}$~cm$^{-2}$. Although the latter are in the minority, namely 23\% of the total number of structures, they contain half of the class 0/I YSOs, proving that they are highly efficient in producing stars. We define a novel statistical metric, the fractality coefficient $\mathcal{F}$, corresponding to the fractal index that an equivalent population of clumps would have if they were generated by an ideal fractal cascade. For NGC$~$2264, over the whole range of spatial scales (1.4-26$~$kAU), this coefficient is globally estimated to be $\mathcal{F}$ = 1.45$\pm$0.12 and its dispersion suggests that the cascade may depend on local physical conditions. However, a single fractal index is not the best fit for the NGC$~$2264 data because the hierarchical cascade starts at a 13$~$kAU characteristic spatial scale.}{Our novel methodology allows us to correlate YSOs with their gaseous environment which displays some degree of hierarchy for spatial scales below 13$~$kAU. We identify hierarchical multi-scale structures, which we associate with a hierarchical fragmentation process, and linear structures, which we associate with a monolithic fragmentation process. Hierarchical structures are observed as the main vectors of star formation. This cascade, which drives efficient star formation, is then suspected of being both hierarchical and rooted by the larger scale gas environment up to 13$~$kAU. We do not see evidence for any hierarchical structural signature of the cloud within the 13-26$~$kAU range, implying that the structure of the cloud does not follow a simple fractal law along the scales but instead might be submitted to a multi-fractal process.}


\keywords{Methods: data analysis - Methods: statistical - Stars: formation - open clusters and associations: individual: NGC$~$2264 - ISM: clouds}

\maketitle

\section{Introduction}

Molecular clouds are large, cold, dense, and substructured gaseous objects that host the formation of young stellar objects (YSOs). Most of these YSOs are not found to be isolated but are rather in close multiple systems \citep{duchene_stellar_2013}, ultra-wide multiple pairs (UWPs, \citealt{joncour_multiplicity_2017}), or are even concentrated in small local stellar groups defined as nested elementary structures (NESTs, \citealt{joncour_multiplicity_2018, gonzalez_s2d2_2021}). These stellar structures cover spatial scales ranging from a few thousand astronomical units (kAU) to a few tens of kAU, suggesting that they originate from a hierarchical fragmentation cascade of the cloud on a larger scale. \\

In order to study the large-scale hierarchy in molecular clouds, \citet{houlahan_recognition_1992} and then \citet{rosolowsky_structural_2008} applied connected tree statistics (dendrograms). This method consists of separating a single image into several pixel-intensity thresholds in order to describe the architecture of a molecular cloud. In these dendrograms, the low-intensity iso-contours of an image are connected to the higher intensity iso-contours. As a result, the dendrogram branches out towards higher intensity levels. By construction, this method extracts a hierarchy in column density, and builds a network whose structure can be described using metrics; for example, the path length between two nodes or the branching of a node. In addition, \citet{pokhrel_hierarchical_2018} studied the hierarchical aspect of sizes using several observations ranging from the cloud scale to the prestellar core scale.  This method has the advantage of being based on a series of independent observations whose associated maps cover a wide range of spatial scales. Consequently, the structural and multi-scale description of the hierarchical cascade of the cloud can be highlighted. However, in contrast to the work of \citet{houlahan_recognition_1992}, the analysis of \citet{pokhrel_hierarchical_2018} does not provide a simple framework to define metrics that would allow the comparison of different regions or of observational results with numerical simulations. Therefore, we propose a novel approach that takes advantage of both the network representation of \citet{houlahan_recognition_1992} and \citet{rosolowsky_structural_2008} and the multi-scale description of the cloud from independent observations \citep{pokhrel_hierarchical_2018}. The latter, coupled with a network representation, would then make it possible to compare, based on network metrics, different star formation regions with numerical simulations. These tools would then supply constraints to star-formation scenarios. In particular, we aim to probe the $\sim$5-30kAU spatial scales in order to look for clues as to the origin of small stellar groups such as UWPs \citep{joncour_multiplicity_2017} and NESTs \citep{joncour_multiplicity_2018}.

In order to test our methodology, we apply it to the molecular cloud NGC$~$2264. Located at 723 pc \citep{cantat-gaudin_gaia_2018} in the Mon OB1 molecular cloud, NGC$~$2264 has a complex star formation history. This cloud has an active star-forming region in its central part, where the majority of class 0/I YSOs have been identified \citep{rapson_spitzer_2014}, as well as two main clusters in the process of forming massive stars \citep{cunningham_submillimeter_2016,nony_mass_2021}. The distribution of YSOs is characterised by an age gradient from the northern part, which contains most of the older stars \citep{rapson_spitzer_2014,venuti_gaia-eso_2018}, to the quiescent southern part \citep{venuti_gaia-eso_2018}. The analysis of this type of cloud could therefore provide clues as to the fragmentation signature of a massive region that may trace the star formation history. \\

This paper is organised as follows. In Section \ref{Section:Multiscale}, we present the multi-scale data of NGC$~$2264 and our graph-theoretic methodology applied to these data. In Section \ref{Section:FractalModel}, we develop a fractal hierarchical cascade model with extended objects before defining a network metric that can describe such a cascade. Using our model and metrics, we study the fractal behaviour of the network of NGC$~$2264. In Section \ref{Section:Discussion}, we discuss the physical origin of the observed properties of the hierarchical cascade. We conclude in Section \ref{Section:Conclusion} with a summary of our results and other possible applications of our methodology.

\section{Multi-scale representation of clumps and YSOs in NGC$~$2264}
\label{Section:Multiscale}

In this section, our goal is to probe the existence of a hierarchical cascade in NGC$~$2264. As we want to investigate this cascade along spatial scales, we use a collection of Herschel maps of NGC$~$2264 obtained at different wavelengths to extract a set of gaseous clumps covering a large range of spatial scales. Using a network-based approach, we then connect the nested clumps and identify disconnected graph components to evaluate the degree of hierarchy in regards to the number of final nodes contained in each of these components.
\subsection{NGC$~$2264 data}
\label{Subsection:Data}

In order to span the $\sim$5-30kAU spatial scales that appear to be the crucial range where hierarchical fragmentation may produce UWPs, NESTs, and clusters of NESTs \citep{joncour_multiplicity_2017, joncour_multiplicity_2018}, we use the Herschel imaging survey of OB YSOs (HOBYS) multi-band observation data \citep{motte_initial_2010} of NGC$~$2264. These data are composed of five emission images at [70, 160, 250, 350, 500]$~\mu$m with respective spatial resolutions of [8.4, 13.5, 18.2, 24.9, 36.3]”, corresponding respectively to spatial scales of $\sim$[6, 10, 13, 18, 26]$~$kAU considering the distance of NGC$~$2264 at 723 pc \citep{cantat-gaudin_gaia_2018}. To extract intensity-peaked clumps in the five Herschel images, we used the multi-scale source and filament extraction method \texttt{getsf} version v210609 \citep{menshchikov_multiscale_2021}. This algorithm applied to an image distinguishes three components which are the sources, the filaments, and the background. As we are only interested here in the extraction of clumps, that is, the sources, we do not take into account the filamentary component which remains embedded in the background component. To extract the sources, the \texttt{getsf} method spatially decomposes the image into several scales in order to give a first identification of the sources and the background. The source images are flattened and small noise fluctuations are removed such that the sources are detected as significant peaks of emission. The geometric and flux properties of the sources are finally measured in the original image {that has been dissociated with} its background. The catalogues produced by \texttt{getsf} contain the spatial characteristics of the ellipsoid sources, in particular the major and minor axes of the ellipses defined as the full width at half maximum (FWHM) of their Gaussian emission profile. The \texttt{getsf} method requires only one input parameter which is the maximum size of the extracted sources, taken here to be about 2.5 times the size of the beam associated with the image. This choice allows us to  not be too selective in the extraction, and to avoid extracting objects that are too extended.
As a result, [231, 202, 126, 92, 46] ellipsoidal clumps are extracted for their respective spatial resolution of [8.4, 13.5, 18.2, 24.9, 36.3]”. This procedure of extraction allows us to monitor the existence of clumps along multiple independent images with different scales and to probe the persistence of the gas structures throughout the scales. 
To complement these five catalogues of clumps, we use a catalogue of YSOs extracted from the Spitzer survey by \citet{rapson_spitzer_2014}. This catalogue originally contains populations of Class III, II, and 0/I YSOs. Only class 0/I YSOs are selected in this work. Indeed, these YSOs are the most likely to be connected to their parental environment, as they are among the youngest; see for example \citet{nony_mass_2021}. As 5\% of class 0/I YSOs are primarily detected by the Multi-Band Imaging Photometer (MIPS), the resolution of which is $\sim$7", and 95\% of these YSOs are also detected with the InfraRed Array Camera (IRAC), the resolution of which is $\sim$2" , we choose 2" to be the associated scale for the class 0/I YSOs. This resolution corresponds to a 1.4$~$kAU spatial scale. As the \texttt{getsf} extraction is performed on a rectangular window of (RA-DEC) corner coordinates [(100.10-10.15), (99.52-9.53), (100.40-8.72), (100.98-9.34)]$^{\circ}$, we only keep the YSOs that are inside this field of view. A total of 87 class 0/I YSOs are kept in this post-selection. Clumps and YSOs are then connected according to the procedure presented in Section \ref{Subsection:Method}. 

A column density image was also computed, which is used in Section \ref{Subsection:SpatialDistribution}. This image at 18.2" resolution is provided by the \texttt{getsf} method \texttt{highres}. This image was derived as follow. Using the [160-500]$\mu m$ emission images, {column} density and temperature images at different resolutions were computed by fitting the spectral energy distribution (SED), assuming optically thin thermal emission. These images were then recombined to produce a high-resolution image {of column density} that contains spatial information from all images at all resolutions. More detailed information about the \texttt{getsf} source extraction method can be found in \citet{menshchikov_multiscale_2021}.



\subsection{Network methodology}
\label{Subsection:Method}

\begin{figure}
    \centering
    \includegraphics[width=9cm]{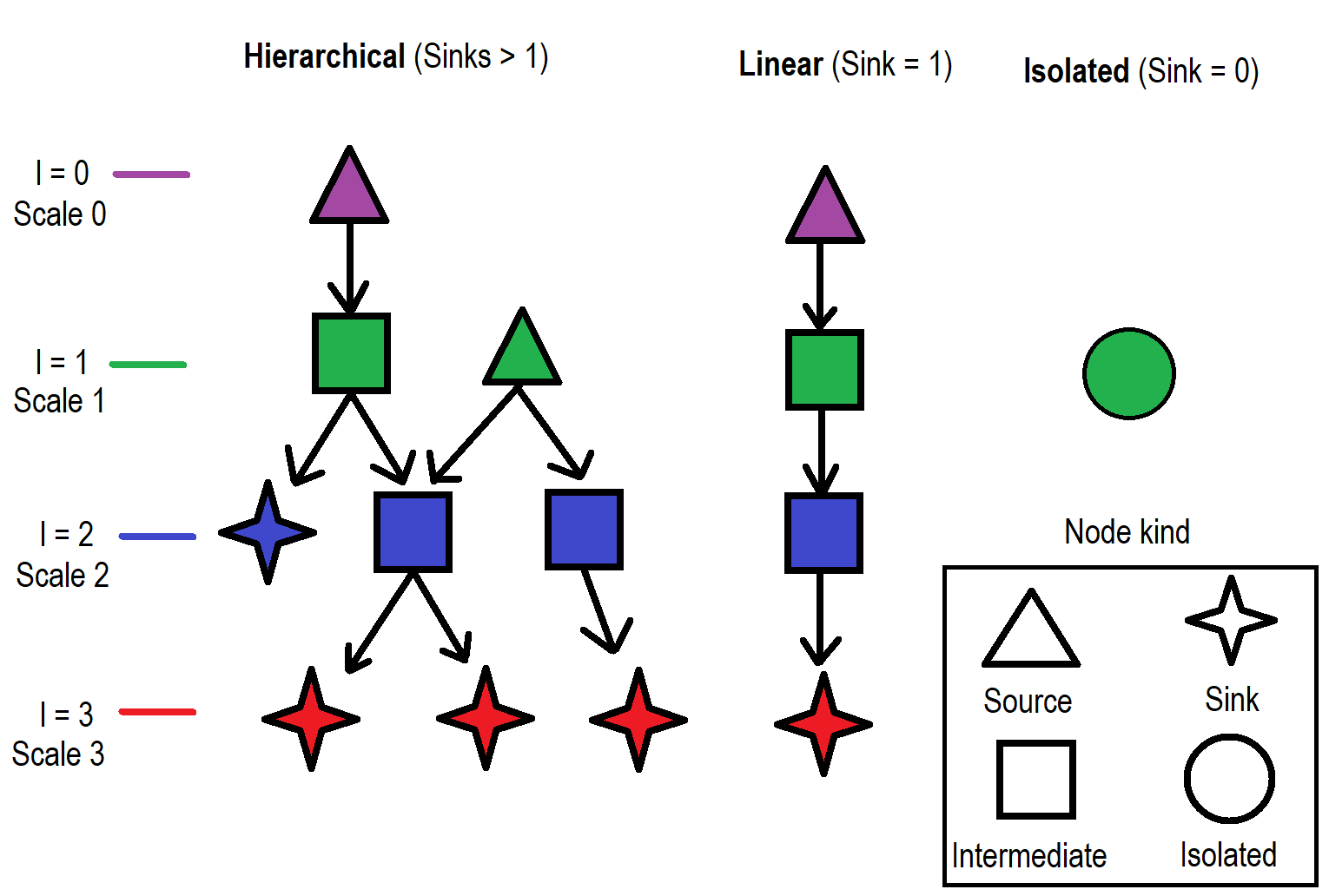}
    \caption{Organisation of a network and definition of nodes. Each colour corresponds to a scale. Each node shape corresponds to a specific kind (source, sink, intermediate, or isolated). Three structures are represented: hierarchical, linear, or isolated. One example of each type of structure is shown.
    }
    \label{Fig:Method}
\end{figure}

In the previous section, we describe how we built catalogues of clumps and YSOs, which we refer to here as substructures for simplicity. Each substructure is associated with a specific spatial scale $s$ which corresponds to the resolution of clump extraction and YSO detection. These substructures can be spatially connected to each other if they overlap, that is, if one substructure is contained within another at a larger scale. The nested substructure is called a `child' while the container is called a `parent'. We define a criterion of connection between two substructures of two different spatial scales. This criterion must take into account the spatial extent of the substructures we consider, both child and parent. We establish the following inclusion criterion: two substructures are connected if at least 75\% \footnote{Here we use the 2$\sigma$ extension of the ellipses. We multiply each axis by $\frac{1}{\sqrt{2 \ln(2)}}$ to get the 2$\sigma$ Gaussian width from the FWHM axis provided by \texttt{getsf} catalogues, and then check the overlap criterion.} of the child total surface is contained inside the extent of the parent. The representation of our data by a connected network appears to be the most adequate description. Indeed, a network is composed of an ensemble of nodes, for which one node contains some data information (in nodal attributes, e.g. size, wavelength, measured flux, etc.). These nodes are connected together with edges that reflect the specific relationship between nodes. In our case, one node describes a substructure which can be a clump extracted from a specific image or a YSO. One edge reflects the inclusion relationship that two substructures might have. Moreover, a network makes it possible to organise the different scales $ s_l $ in different levels $ l $ as shown in Figure \ref{Fig:Method}. Each of our six catalogues is associated with an $l $ level so that a total of six levels can be defined. The edge created between two nodes is oriented from the parent (high-scale) to the child (low-scale) such that an oriented network is constructed from $l = 0$ to $l = L-1$, where L is the total number of levels. In this work, $L = 6$ with $s_0 = 26~kAU$ and $s_5 = 1.4~kAU$. The number of substructures at each level $l$ can be written as: $\{N_0, N_1, ..., N_{L-1}\}$. With this method of connection between substructures and between scales, the resulting connected network contains several subnetworks that are independent of each other. These subnetworks are disconnected from their counterparts; they are components of the main network. From a physical point of view, this organisation reflects the spatial correlation a parent has with its children. Indeed, a single component describes a whole complex of connected substructures that are all affiliated. On the contrary, two different components describe two spatially unaffiliated complexes of substructures and two independent structures are defined. In the remainder of this paper, we use the term `structure' to refer to these network components. Inside a structure, there are nodes connected to each other. These nodes are labelled as follows:
 
 \begin{itemize}
    \item Graph-sinks: nodes without children that inherit from at least one parent.
    \item Graph-sources: nodes without a parent that breed at least one child.
    \item Intermediates: nodes with child(ren) and parent(s).
    \item Isolated: nodes without a child or a parent.
\end{itemize}
 
 If we write as $\{N_0, N_1, ..., N_{L-1}\}$ the number of substructures contained in the structures shown in Figure \ref{Fig:Method} at all levels $l$, we define the following structures types:
 
 \begin{itemize}
    \item Hierarchical: Structure with multiple substructures in at least one level; for example $\{1,2,3,3\}$ substructures at level $l$.
    \item Linear: Structure with at most one single substructure at all levels; for example $\{1,1,1,1\}$ substructures at level $l$.
    \item Isolated: Structure that is an isolated node; for example $\{0,1,0,0\}$ substructure at level $l.$
\end{itemize}
 
We can then differentiate between non-persistent structures (isolated), structures that physically only represent a single object (linear), and structures that show some degree of multiplicity and clustering between substructures (hierarchical). We use this procedure to construct a network for NGC$~$2264  in the following section (Section \ref{Subsection:StructuresInNGC}).

\subsection{Extracted structures in NGC$~$2264}
\label{Subsection:StructuresInNGC}

\begin{figure}[!h]
\centering
    \includegraphics[width=9cm]{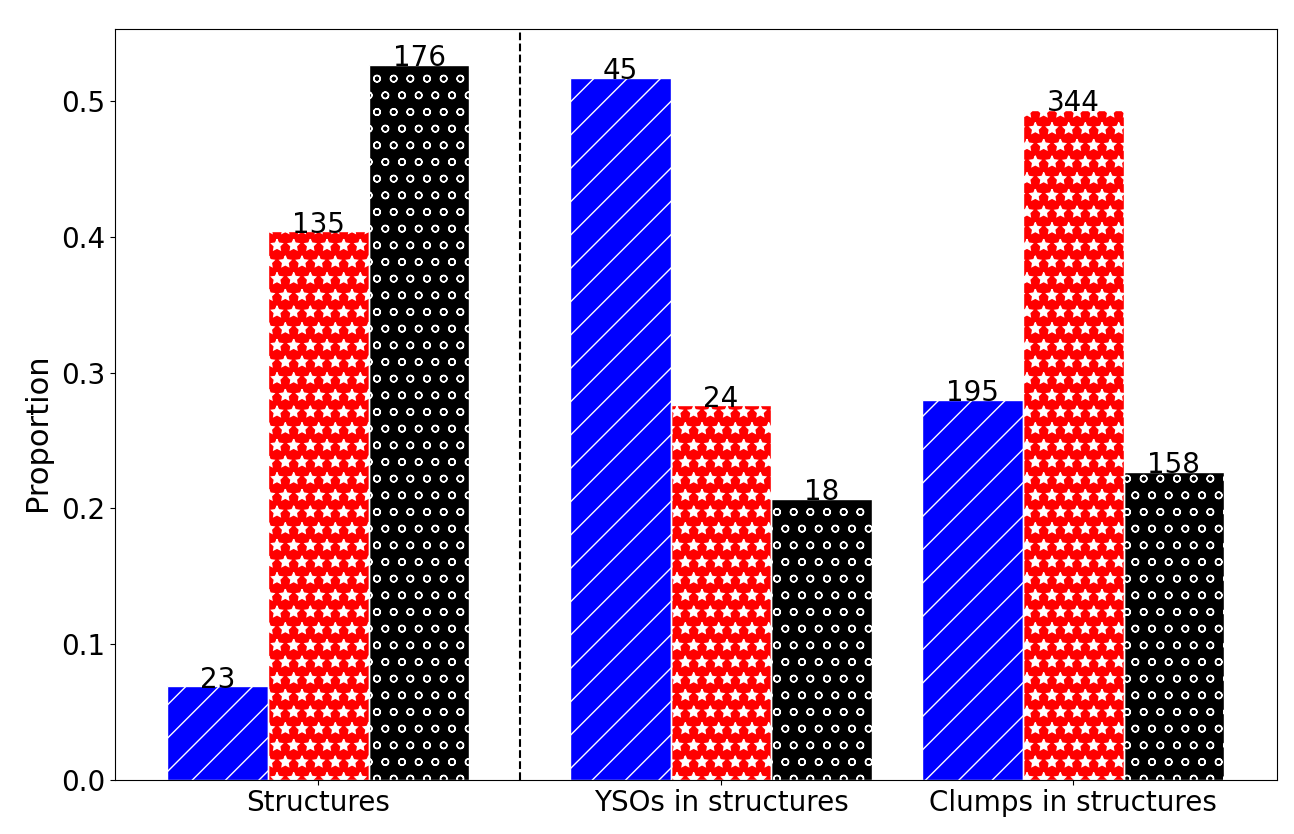}
    \caption{\textit{Left}: Statistics of extracted structures. Precise numbers are shown above the bars. Hierarchical structures {are represented with blue bars (line hatching)}, linear with red {bars (star hatching), and isolated with black bars (circle hatching}). \textit{Right}: Statistics for each type of substructure (YSOs or clumps) inside each type of structure.}
    \label{Figure:piechart}
\end{figure}

\begin{figure*}[!h]
\sidecaption
\includegraphics[width=12cm]{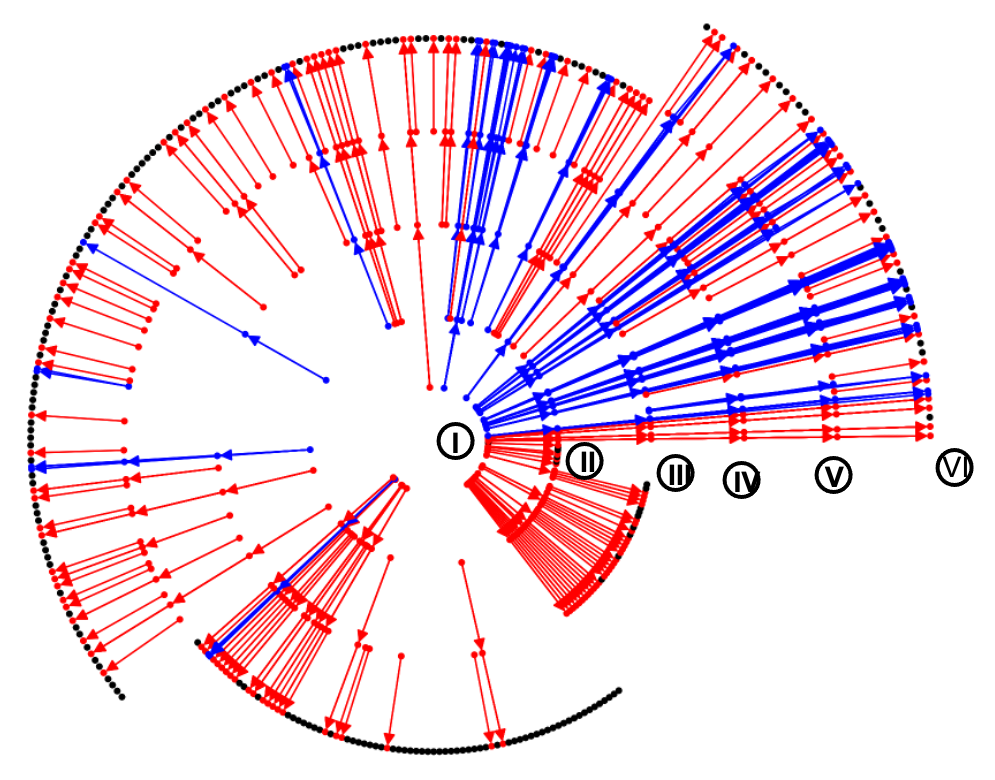}
\caption{Network representation of the data. Each substructure is represented as a node with respect to its specific scale, associated with a ring. Rings from I to VI represent respectively [26, 18, 13, 10, 6]$~$kAU and class 0/I YSOs at 1.4$~$kAU such that inner rings represent the larger substructures and the scale is decreasing towards the outer rings. The angular position {indicates} the membership of substructures to a structure. The cascade of one structure is processed in the same direction towards the exterior of the wheel. Blue, red, and black colours represent hierarchical, linear, and isolated structures, respectively.}
\label{Figure:wheel}
\end{figure*}

In NGC$~$2264, the network built by the procedure described in Section \ref{Subsection:Method} is composed of six levels of scale: five of them contain clumps at their associated resolution and one contains class 0/I YSOs. The network involves a total of 334 structures in which 23 are hierarchical, 135 are linear, and 176 are isolated (see Table \ref{table:extraction} and left panel of Figure \ref{Figure:piechart}). Inside the same class of structure, one can observe the different physical nature of graph-sinks (see Figure \ref{Figure:wheel}) as they can be associated either to YSOs or clumps. In fact, some structures do not cascade down to the last  possible level, but stop before that. A structure that contains a YSO might be in a more evolved stage of collapse since star formation actually happened. The diversity in the nature of the graph-sinks {inside} hierarchical structure can also indicate the global evolutionary stage. Indeed, if a hierarchical structure contains YSOs as graph-sinks only, it is more likely to have ended its star formation process than a structure hosting sibling YSOs and clumps. With this representation, the existence of YSOs inside a structure can be readily evaluated along with their multiplicity.

Hierarchical structures contain 52\% of the total YSO population and 28\% of the total gaseous substructures (see Figure \ref{Figure:piechart}). As a majority of YSOs are forming inside hierarchical structures despite their under-representation in the cloud ($\sim$7\%), these structures are not irrelevant when it comes to the formation of young stars. Linear structures contain 50\% of the population of whole clumps  but host only 28\% of the total YSO population in the cloud. To further evaluate the role of hierarchy in star formation, we can assess the average production of YSOs and the proportion of graph-sinks they represent. The hierarchical structures produce on average 2.0 YSOs while the linear structures produce on average 0.2 YSOs. We would expect the linear structures to produce a maximum of one single YSO, which means that most of the linear structures do not cascade down to the stage of forming a young star (see Figure \ref{Figure:wheel}), or at least not yet. In addition, the ratio (YSOs: graph-sinks) contained in a type of structure indicates {the actual proportion of YSOs found} at the end of the cascade. We find that linear structures (8: 45 ratio) are mainly composed of gaseous graph-sinks whereas hierarchical structures (45: 94 ratio) show a more dominant presence of YSOs. 

Consequently, hierarchical structures dominate the total production of YSOs in the cloud and cascade more towards lower levels compared with linear structures. Therefore, hierarchical structures seem to be either more evolved than their linear counterparts, or are the main vectors for the formation of YSOs; this latter case would mean that hierarchical structures are closely linked to the formation of stellar groups.

\begin{table}[!h]
\caption{Number of structures of each class inside NGC$~$2264 with the numbers of clumps and YSOs inside each of them. An X indicates the absence of graph-sinks and graph-sources within isolated structures as these structures do not contain them by definition.}
\label{table:extraction}
\centering 
\begin{tabular}{c c || c c c | c} 
    &   & \multicolumn{3}{c|}{Structures}           & \\ 
\cline{3-5}
    &   & Hierarchical   & Linear   & Isolated     & Total  \\
\hline
\multicolumn{2}{c||}{Number}    & 23 & 135 & 176 & 334 \\
\hline \hline
\multicolumn{2}{c||}{YSOs included}          & 45 & 24 & 18 & 87  \\ 
\hline
\multirow{6}{3em}{Clumps included} 
 & 70$~\mu$m  & 69 & 77 & 85 & 231 \\
 & 160$~\mu$m & 52 & 91 & 59 & 202 \\
 & 250$~\mu$m & 33 & 84 & 9 & 126\\
 & 350$~\mu$m & 25 & 63 & 4 & 92 \\
 & 500$~\mu$m & 16 & 29 & 1 & 46 \\
 & Total & 195 & 344 & 158 & 697 \\
 \hline
\multicolumn{2}{c||}{YSOs + clumps} & 240 & 368 & 176 & 784 \\
\hline
\hline
\multicolumn{2}{c||}{Graph-sinks} & 94 & 135 & X & 229\\
\multicolumn{2}{c||}{Graph-sources} & 34 & 135 & X & 169 \\
\end{tabular}
\end{table}

\subsection{Spatial distribution}
\label{Subsection:SpatialDistribution}

\begin{figure*}[!h]
\centering
\includegraphics[width=12cm]{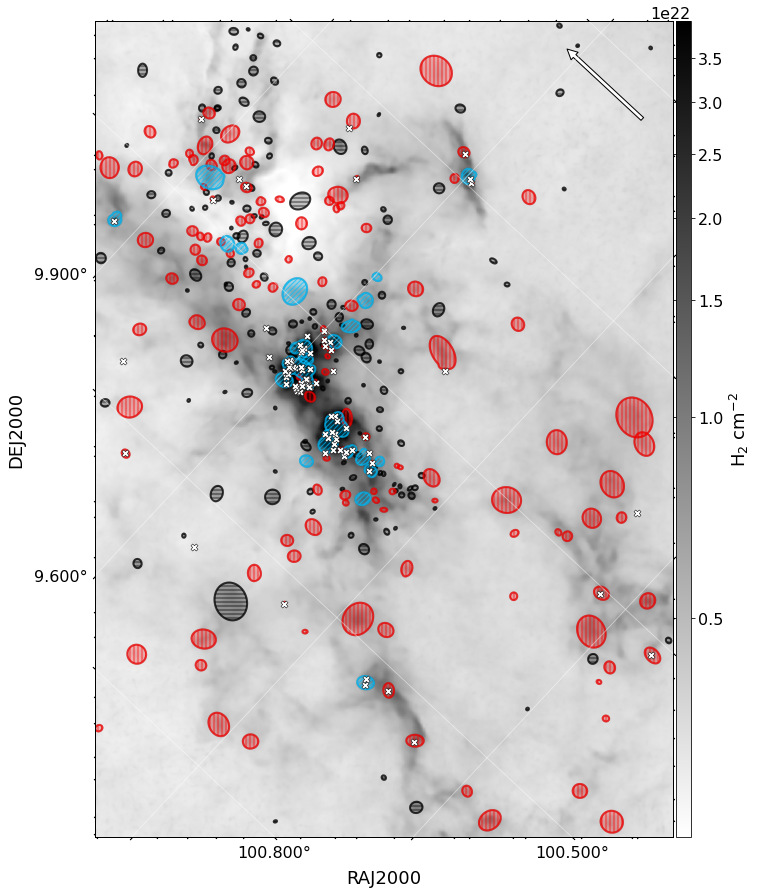}
\caption{Column density image of NGC$~$2264 at 18.2" resolution. Extracted multi-scale structures are superimposed as coloured imprints with the class 0/I YSOs (white crosses). Cyan (diagonal hatch), red (vertical hatch), and black (horizontal hatch) ellipses locate hierarchical, linear, and isolated structures, respectively. The white arrow indicates the north.}
\label{Figure:map_regions}
\end{figure*}

\begin{figure}[!h]
\centering
\includegraphics[width=9cm]{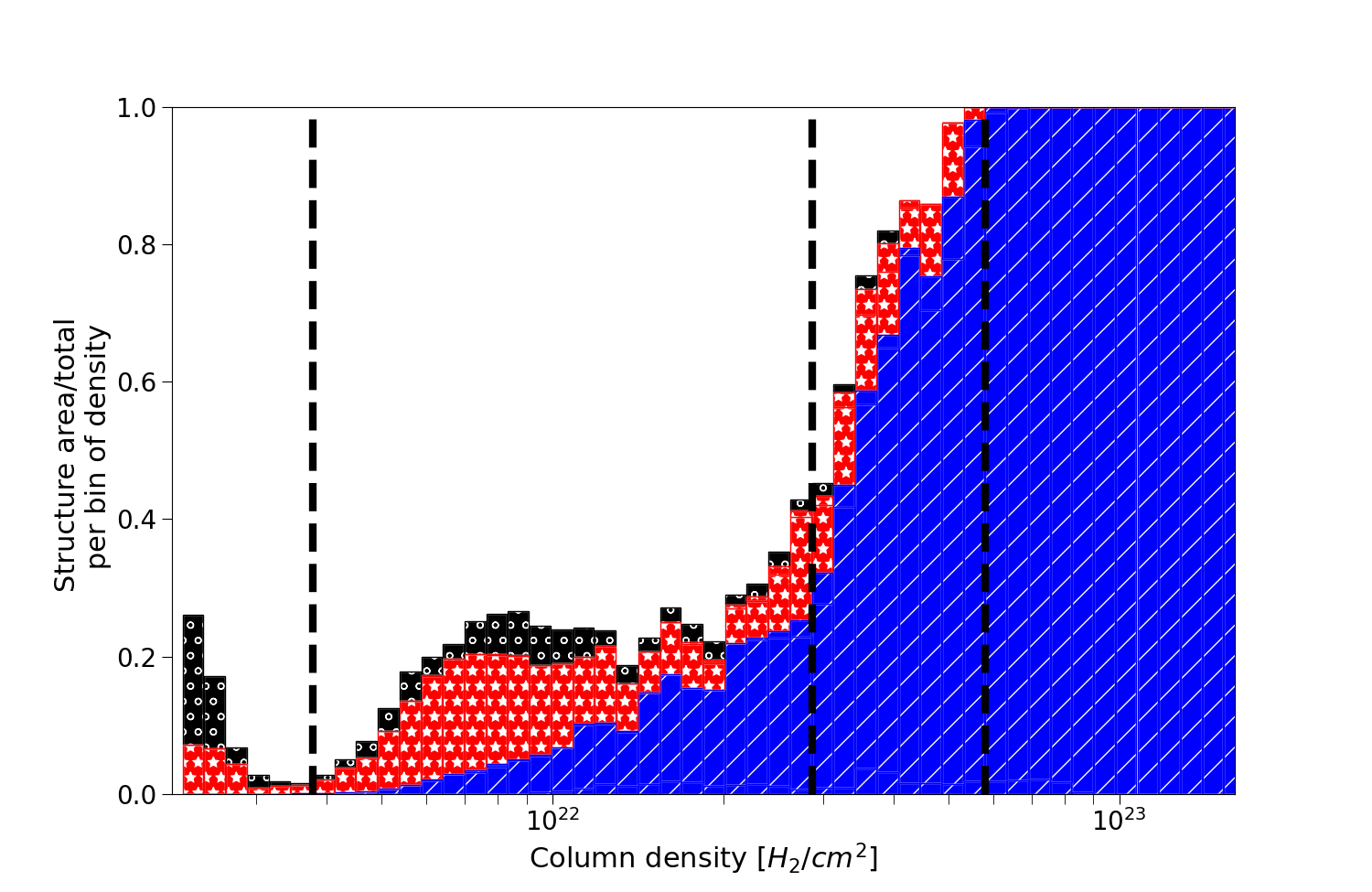}
\caption{Stacked histogram of the surface covered by structures per bin of surface density, normalised by the total area covered by a bin of surface density. White bars are unassigned pixels, red (stars) are linear structures, blue (lines) are hierarchical, and black (circles) are isolated. Vertical lines represent threshold for the proportion of area covered by hierarchical structures, from left to right: 0\%, 40\%, 100\%.}
\label{Figure:hist_density}
\end{figure}

Another major difference between the types of structures we extracted despite their YSO multiplicity is their location in the cloud with respect to the column density background $\Sigma$. In NGC$~$2264, structures appear to be partitioned with graph-sources of all scales. The hierarchical structures are mainly localised in the two central hub regions (see Figure \ref{Figure:map_regions}), while the linear structures are mostly localised in the north and south filaments, and on the eastern region. Isolated structures are distributed everywhere and do not seem to have a preferential location. In order to investigate the correlation between structure type and local column density, we associated each class of structure to the number of pixels of the column density image they cover. Then,  inside a bin of column density $B_\Sigma$, we evaluated the proportion of pixels contained in a structure compared to the total number of pixels of $B_\Sigma$. In other words, we evaluate the proportion of area covered by a type of structure inside each bin of column density. We find that hierarchical structures occupy more than 95\% of the map area with column density $\boldsymbol{N_{\rm H_2}} > 6 \times 10^{22} cm^{-2}$ (see Figure \ref{Figure:hist_density}). Hierarchical structures start being dominant for density $\boldsymbol{N_{\rm H_2}} > 3 \times 10^{22} cm^{-2}$ with an area occupation $> 40\%$ while below $\boldsymbol{N_{\rm H_2}} \sim 3 \times 10^{22} cm^{-2}$ all structures coexist. This behaviour is consistent with the definition of hubs and ridges, which are high-density (and therefore high-column density) cloud structures forming clusters
of stars with a high star-formation rate (see e.g. \citealt{louvet_w43-mm1_2014, motte_high-mass_2018}). 
In addition, the linear structures show a peak of presence at $N_{\rm H_2} \approx 6 \times 10^{21} cm^{-2}$. This peak could indicate that local physical conditions govern the successive emergence of linear and hierarchical modes in the cloud. As in this work we only focus on the  analysis of hierarchical structure, this effect is not investigated further here. It is also important to consider that the cascade might appear linear but may become hierarchical at higher resolution, which we do not probe here. This effect might convert some of the pixels associated to linear structures into hierarchical structures, and thus increase the total hierarchy in the cloud. 

\section{Towards a fractal model of the hierarchical cascade}
\label{Section:FractalModel}

In  Section \ref{Section:Multiscale}, we identify  different types of multi-scale structures in NGC$~$2264. In this section, we focus on the hierarchical structures we highlighted. This hierarchical cascade is occurring inside the densest region of NGC$~$2264, and dominates the star production in the cloud. However, the characteristics of this cascade throughout the scales has not
yet been investigated. In this section we aim to evaluate the scale-free property of this cascade. For this purpose, we compare the hierarchical cascade of NGC$~$2264  with a fractal cascade model. To make a proper comparison, our model displays the spatial properties of substructures: these are spatially extended as ellipsoids and belong to a specific scale. We also define a network metric to ease the comparison between the cascade of  NGC$~$2264  and the fractal model.

\subsection{Geometrical fractal model generation}
\label{Subsection:ModelGeneration}


\subsubsection{Parameters and bases}
\label{Subsubsection:Parameters}

The model proposed here consists of generating a population of extended ellipsoidal clumps that subdivide into nested and smaller ellipsoids along the spatial scale at different levels $l$. In order to reproduce the same network architecture that we build for the data, we define multiple levels $l$ associated to scales $s_l$. As the model is scale-free, the only important quantity is the scaling ratio $r_{l \rightarrow l+1} > 1$ between two spatial scales $s_l$ and $s_{l+1}$ with $s_l > s_{l+1}$:
\begin{equation}
    r_{l \rightarrow l+1} =\frac{s_l}{s_{l+1}}
    \label{eq:r}
.\end{equation}

The value of the initial scale $s_0$ is intrinsically irrelevant as the process we want to model is scale-free; however, it allows us to define all the levels in the network and to assign a physical dimension to the substructures we model. For example, the scale $s_1$ is defined with respect to $s_0$ with the scaling ratio $r_{0 \rightarrow 1}$. These substructures subdivide following a fractal law of constant fractal index $N_{[r_0]}$ such that $N_{[r_0]}$ children are generated each time the scale of the parental substructure is reduced by a factor $r_0$. If the scale is reduced by any factor $r_{m \rightarrow l}$ between two levels $m$ and $l$, we can always define $x$ such that $r_{m \rightarrow l} = {r_0}^x$ where the exponent $x$ corresponds to the number of times the scale has been reduced by $r_0$ between levels $m$ and $l$. The exponent $x$ can be written as:

\begin{equation}
    x = \frac{\ln r_{m \rightarrow l}}{\ln r_0}
    \label{eq:xl}
.\end{equation}

The number of substructures $N_{m \rightarrow l}$ produced from the level $m$ to the level $l$ is then ${N_{[r_0]}}^x$, which leads to

\begin{equation}
    \ln{N_{m \rightarrow l}} = \frac{\ln r_{m \rightarrow l} }{\ln(r_0)} \times \ln{N_{[r_0]}} 
    \label{eq:Neta1}
.\end{equation}

In particular, the total number of substructures $N_l$ produced by a level $s_0$ graph-source  at the level $l$ is given by
\begin{equation}
    \ln{N_l} = \frac{\ln \left( \frac{s_0}{s_l} \right ) }{\ln(r_0)} \times \ln{N_{[r_0]}} 
    \label{eq:Neta}
.\end{equation}

As $N_{[r_0]}$ is defined with respect to an arbitrary scaling ratio $r_0$, the scaling ratio $r_{l \rightarrow l+1}$ we request in the model does not affect the total multiplicity at the scale $s_l$. Indeed, with these definitions, the fractal index allows a parental substructure to subdivide into the right number of pieces irrespective of the value of $r_{l \rightarrow l+1}$. For example, a parental node in level $l$ with fractal index $N_{[2]}=2$ produces 2 substructures for $r_{l \rightarrow l+1}  = 2$; and $2^2 = 4$ substructures for $r_{l \rightarrow l+1} = 4,$ which is consistent. Without this scaling ratio of reference $r_0$ in the definition of the fractal index, the production of substructures would be the same for two different $r_{l \rightarrow l+1}$, which can lead to different multiplicity, and therefore to inconsistencies when we stop at the same scale whether we take two, three, or more levels to reach it.

\subsubsection{Generative subdivision process}
\label{Subsubsection:FragmentationProcess}

Once the two parameters $N_{[r_0]}$ and $r_{l \rightarrow l+1}$ of the model are defined, we can establish a procedure to generate extended substructures. To start this generative process, an initial population of $N_{ini}$ {graph-sources} is defined at the initial scale $s_0$. Each graph-source subdivides into $N_1$ pieces of size $s_1 < s_0$. Sizes are determined by the scaling ratio $r_{0 \rightarrow 1}$ (see equation \ref{eq:r}) requested in the process. $N_1$ is determined by equation \ref{eq:Neta} with $l = 1$. Each sibling placement inside a graph-source takes into account the spatial extent of substructures (see Section \ref{Subsection:Selection}). This procedure is repeated for all scales $s_l \rightarrow s_{l+1}$ until the last scale $s_L$ is reached. As $N_l$ represents the total number of substructures produced by a {graph-source} at the level $l$, we redistribute these $N_l$ substructures equally into their $N_{l-1}$ parents, which will contain $N = N_l/N_{l-1}$ substructures each. For the case in which $N$ is a non-integer number, the selection of the `real' $\widetilde{N}$ is done randomly using a binomial rule between $\lfloor{N}\rfloor$ and $\lceil{N}\rceil$ with respective probability $1-\{N\}$ and $\{N\}$; where $\lfloor{x}\rfloor$ and $\lceil{x}\rceil$ denote the floor and ceiling function of $x,$
respectively, and $\{x\}$ denotes its fractional part. It is not excluded that two graph-sources share a child in common if their extent overlaps; in this instance, they would be part of the same structure. Consequently, a structure generated by our model possesses on average $N_{ini} \times N_l$ substructures at the level $l$, but individual structures may not host exactly $N_{ini} \times N_l$ substructures if $N_{ini} \times N_l$ is a non-integer.


\subsection{Rules of substructure selection}
\label{Subsection:Selection}

\subsubsection{Geometrical parameter sampling}

Because our model generates extended objects, it is necessary to establish rules in order to synthesize a coherent population both in terms of spatial coverage inside the same level $l$, and in terms of inheritance. Each substructure is defined by its centroid coordinates $(x,y)$; by its major and minor axis $(a,b),$ respectively; and by its position angle $\theta$ in 2D space. The coordinates of the graph-sources are sampled uniformly in a squared window while the coordinates of the children are sampled uniformly inside their parental ellipsoids (more details in Sections \ref{Subsubsection:Rule2} and \ref{Subsubsection:Rule3}). The orientation is sampled uniformly between 0 and $\pi$. The major axis $a$ is sampled in a normal distribution $\mathcal{N}(s_l,\frac{s_l}{10})$\footnote{The probability of getting a negative value is given by the cumulative function $P(x<0) = \frac{1}{2} \left ( 1+erf(-\frac{0 - \mu}{\sigma \sqrt{2}}) \right )$, where $erf$ represents the error function. This probability is independent of $s_l$ because $\mu = s_l$ and $\sigma = \frac{s_l}{10}$ in our case. As $P(X<0)<1.10^{-16}$, the probability of sampling a negative value is unlikely.} for which the mean value $s_l$ is the scale of the substructure inside the level $l$, and the standard deviation is equivalent to 10\% of its typical scale to allow small fluctuations. The minor axis is taken from the value of the major axis and the aspect ratio which is sampled uniformly between 0.5 and 1 as \texttt{getsf} extraction rejects clumps with aspect ratios $<$ 0.5.

\subsubsection{Rule\#1: Minimal distance between substructures of an $l$ level}
\label{Subsubsection:Rule1}

Two substructures that belong to the same level $l$ cannot be closer (taking the centroid distance) than their typical scale, or they would blend. Their spatial extent can only overlap up to a certain limit. In this model, the typical scale $s_l$ of generated substructures is {taken to be} 2$\sigma$ {of the} extent of a real substructure. However, in observational data, the typical scale of substructure, which can be associated to the image beam, is defined {with respect to} their FWHM. In our model, two substructures of scale $s_l$ cannot be closer than $\sqrt{2\ln(2)} \times s_l$. The factor $\sqrt{2\ln(2)}$ is the scaling factor between FWHM and 2$\sigma$ such that the scale defined in the model is comparable to the scale one can define in real data. This leads to \textit{Rule\#1}: the minimal distance between the centroids of two substructures inside the level $l$ is $\sqrt{2\ln(2)} \times s_l$. 

\subsubsection{Rule\#2: Child coverage inside its parent}
\label{Subsubsection:Rule2}

In Section \ref{Subsection:Method}, the inheritance property of the network is defined as the inclusion relationship between a child and its parent. The two are connected if 75\% of the child spatial extent is contained inside its parent (see Fig \ref{Figure:selection}). This relationship needs to be kept when a child is placed inside a parent during the generative procedure. In polar coordinates $[R; \phi]$, the equation of the border of an ellipse of half major axis $a$ and half minor axis $b$ is:

\begin{equation}
        R_{\rm max} = a \sqrt{\cos^2{\phi} + \left(\frac{b}{a}\right)^2 \sin^2{\phi}}
.\end{equation}
The child centroid coordinate inside such an ellipse is sampled in:
\begin{equation}
    \left\{
    \begin{aligned}
        R & \in \mathopen[0,R_{\rm max}\mathclose] \\
        \phi & \in \mathopen[0,2\pi\mathclose]
    \end{aligned}
    \right.
.\end{equation}
A threshold limit is then imposed to verify that the proportion of the area of the child that lies within that of its parent actually exceeds 75\%. This defines \textit{Rule\#2}: a parent needs to cover at least 75\% of the area of its child for this child to actually exist and be placed. This threshold is purely arbitrary and 75\% is chosen because this is the tolerance requested to define our structures in the  data for NGC$~$2264.

\subsubsection{Rule\#3: Room for siblings}
\label{Subsubsection:Rule3}

\begin{figure}
\resizebox{7cm}{!}{\includegraphics{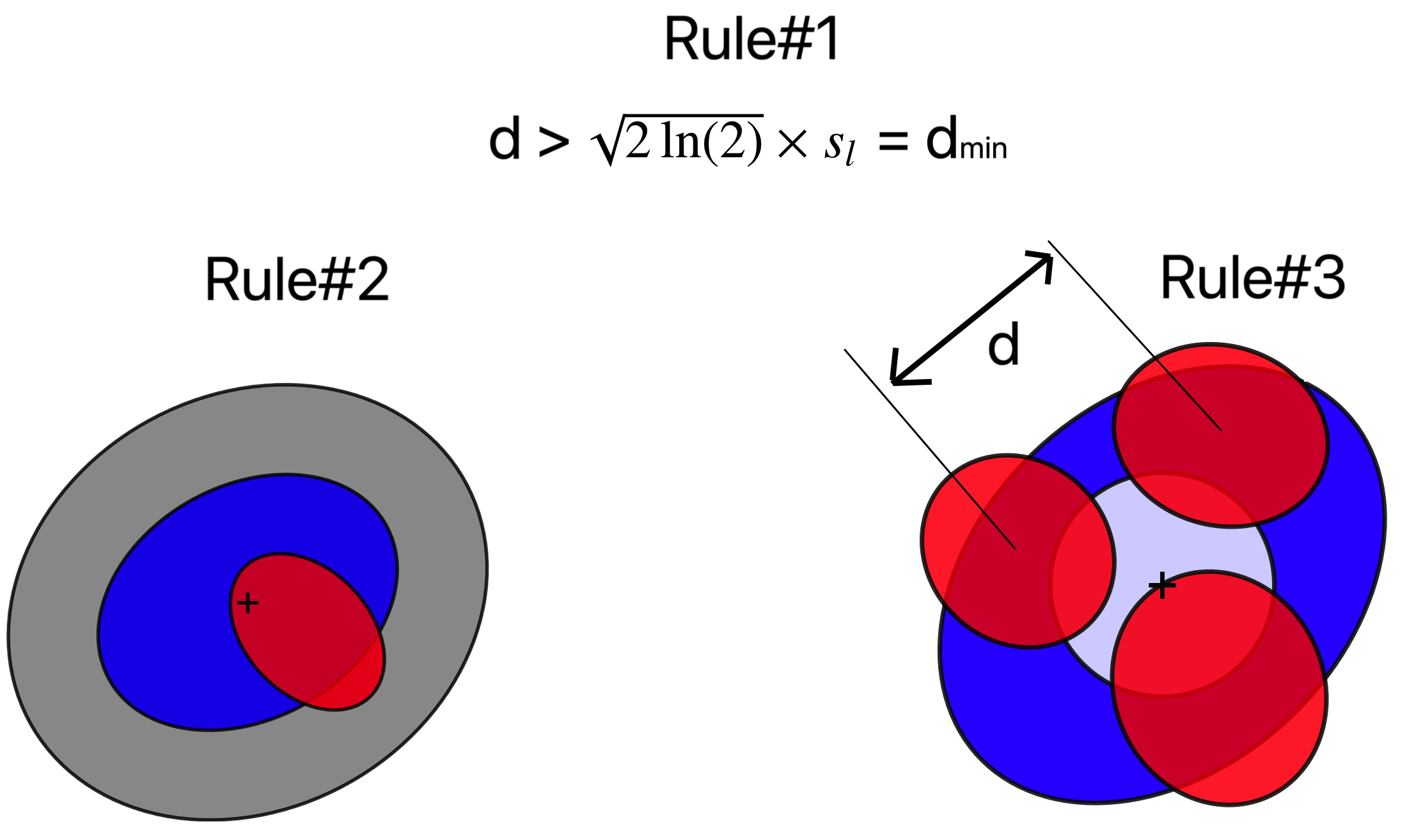}}
\caption{Rules of selection for the model. \textit{Rule\#1}: The distance $d$ between the centroids of two substructures  of the same level at scale $s_l$ needs to exceed the indicated tolerance. \textit{Rule\#2}: As the child (red) lies within its parent (blue), its extent can go beyond the grey area as long as 75\% of its area remains inside its progenitor. \textit{Rule\#3}: Child (red) centroid lies within its parent (blue), but avoids the forbidden area at the centre (grey). The minimal separation $d_{min}$ between the centroids of siblings  is seen as the diameter of the forbidden area.}
\label{Figure:selection}
\end{figure}

During the generation of hierarchical structures, it is expected that more than one child need to be placed inside a parent. When adding siblings, a geometrical constraint can appear if the first selected child lies in the middle of its parent. A child placed in the central area may prevent its siblings from being placed as its presence drastically reduces the available space. If this happens, the parental clump may be overcrowded and there could be insufficient space available to actually select a sibling according to Rule\#1 and Rule\#2. To reduce this geometrical constraint, the inner area of the parent is forbidden for the centroid selection of a child. By adding this forbidden area, the probability of choosing a child that prevents the placement of siblings is reduced, relaxing the constraint. The full extent of this inner region is taken as the minimal separation between two clumps (consequence of Rule\#1), and is supposed to be circular (isotropic because the distribution of orientation $\theta$ is uniform). Therefore, the radius of the inner region is half of its full extent (see Fig. \ref{Figure:selection}). In case of multiple children, centroid coordinates are selected according to \textit{Rule\#3} which alters the original selection as follows:

\begin{equation}
    \left\{
    \begin{aligned}
        R & \in \left[\frac{\sqrt{2\ln(2)} \times s_l}{2},R_{\rm max}\right] \\
        \phi & \in \mathopen[0,2\pi\mathclose]
    \end{aligned}
    \right.
.\end{equation}

We note that if only one child is desired, this area becomes useless and the child is able to be placed anywhere inside its parental area, provided that Rule\#2 is fulfilled. Consequently, our model generates a population of substructures and takes into account their spatial extent. The resulting network mimics those that can be derived from a real observational data set and can be tested.

\subsection{Metric to describe a fractal hierarchical cascade}
\label{Subsection:Metrics}

\subsubsection{A measure of the network fractality}
\label{Subsubsection:Fractality}

Our geometrical fractal model generates a population of extended substructures. These substructures can form hierarchical structures (for $N_{[r_0]} > 1$) or linear structures (for $N_{[r_0]} = 1$). As we focus on hierarchical structures, we consider that the population of structures generated by our model is hierarchical. In order to characterise the hierarchical cascade in terms of multiplicity, we define a network fractality coefficient $\mathcal{F}$. This coefficient is a measure of the fractal index that would best describe a network (e.g a hierarchical structure). Computing this coefficient for a structure generated by our model returns the input free parameter $N_{[r_0]}$. Therefore, by measuring $\mathcal{F}$ in the data, we aim to estimate the parameter $N_{[r_0]}$ that would best fit our model. To compute this coefficient $\mathcal{F}$ in the data, we first measure the total number of nodes (i.e. substructures) $V_0$ of a network $G_0$ that possesses $L$ levels of scales $s_l$. We then compare this number of nodes with the expected number of nodes produced by an ideal fractal network using equation \ref{eq:Neta}. Consider an ideal fractal network $G$ that subdivides into $N_{[r_0]}$ pieces for each scale reduction of an arbitrary factor $r_0$. This network of reference contains the same levels as $G_0$. On one hand, $G_0$ possesses a total number of $V_0$ nodes. On the other hand, the total number of nodes $V$ possessed by  $G$ is the sum over the levels of all nodes contained in the level $l$. To lighten the notation, we consider here $N = N_{[r_0]}$. The number of nodes at a level $l$ is, according to equation \ref{eq:Neta}, $N^{x_l}$ where $x_l = \frac{\ln (\nicefrac{s_0}{s_l})}{\ln (r_0)}$. The total number of nodes belonging to 
$G$  is
\begin{equation}
    \begin{cases}
    V = N^{x_0} + N^{x_1} + ... + N^{x_L} = \sum_{l=0}^{L} N^{x_l} \\
    \\
    x_l = \dfrac{\ln \left (s_0 / s_l \right)}{\ln (r_0)}
    \end{cases}\,.
\end{equation}
Because both networks need to be equivalent in terms of the number of nodes for comparison, we have $V_0 = V$. When this condition is fulfilled, the measure of the fractality $\mathcal{F}$ of  $G_0$ is equivalent to the fractal index $N$ by definition. Hence, $\mathcal{F} = N$ when $V_0 = V$. The previous equation can therefore be written as

\begin{equation}
    \begin{cases}
    V_0 - \sum\limits_{l=0}^{L-1} \mathcal{F}^{x_l} = 0 \\
    \\
    x_l = \dfrac{\ln \left (s_0 / s_l \right)}{\ln (r_0)}
    \end{cases}\,.
    \label{eq:fractality}
\end{equation}

Consequently, $\mathcal{F}$ is the root solution of the previous equation, and can be computed numerically. With the definition of the exponent $x_l$, $\mathcal{F}$ is independent of the set of scales $s_l$ available. Indeed, a different set of scales would have no impact on the theoretical value of $\mathcal{F}$ because it is defined with respect to a fixed, arbitrary scaling ratio $r_0$. However, some of the hierarchical structures in the data for NGC$~$2264  contain multiple graph-sources, whereas Equation \ref{eq:fractality} accounts for networks with a single graph-source. For each graph-source $i$ of a structure, we compute a fractality $\mathcal{F}_i$ that is computed from the subnetwork described by all the nodes that are directly affiliated to this graph-source $i$. For example, consider a hierarchical structure with a graph-source $i = 0$ that is linked with nodes labelled $A$, $B$, and $C$. Another graph-source $i = 1$ is affiliated to nodes $C$, $D,$ and $E$. Then, $\mathcal{F}_0$ is computed from nodes $A$, $B$, and $C$ regardless of $D$ and $E,$ even though these latter are part of the structure. Similarly, $\mathcal{F}_1$ is computed from nodes $C$, $D$, and $E$. Hence, the fractality of the whole structure is defined as the average of these individual fractality coefficients over all the graph-sources:

\begin{equation}
\mathcal{F} = ~ <\mathcal{F}_i>_{\rm sources} ~~ = ~~  \frac{ \sum\limits_{i = 0}^{N_{\rm sources}} \mathcal{F}_i }{N_{\rm sources}}
.\end{equation}

The measurement of $\mathcal{F}$ can be performed for any network built under our graph-theoretic representation, whether data comes from the model or real data. From this measurement, we can assess the number of subdivisions one equivalent fractal network would have with the same number of levels. We can evaluate the production of substructures scale by scale and compare it with an equivalent fractal network using our fractal model.


\subsubsection{{Structure} productivity at each scale}
\label{Subsection:Productivity}

The productivity of {structures} evaluates, at each scale $s_l$ of a structure, the number of substructures that appears with respect to the number of graph-sources $N_{\rm sources}$ in this structure. This is the ratio between the number of substructures at a scale $s_l$, and the number of graph-sources of the structure. Normalising by $N_{\rm sources}$ is interesting because if a structure contains multiple graph-sources, the total number of substructures (assuming an ideal fractal cascade) would be multiplied by $N_{\rm sources}$, and the production rate might be biased. For a fractal process of index $N_{[r_0]}$, Equation \ref{eq:Neta} gives the relationship between the number of substructures $N_l$ produced at the scale $s_l$ and $N_{[r_0]}$. By dividing the equation by $N_{\rm sources}$:

\begin{equation}
    \frac{N_l}{N_{\rm sources}} = \frac{N_{[r_0]}^{\nicefrac{\ln \left( \frac{s_0}{s_l} \right ) }{\ln r_0}}}{N_{\rm sources}}
.\end{equation} \\

This leads to 
\begin{equation}
    \ln \left( \frac{N_l}{N_{\rm sources}} \right) =  \frac{\ln \left( \frac{s_0}{s_{l}} \right ) }{\ln(r_0)} \times \ln{N_{[r_0]}} - \ln N_{\rm sources}
.\end{equation}
Letting $Y = \ln \left( \frac{N_l}{N_{\rm sources}} \right) $ and $X = \ln \left( \frac{s_0}{s_{l}} \right ) $ , the relation can be written as

\begin{equation}
    Y  =  \frac{\ln{N_{[r_0]}} }{\ln(r_0)} \times X  - \ln N_{\rm sources}
    \label{eq:slope}
.\end{equation}

The productivity of an ideal fractal cascade should be described by a power law in which the slope in a log--log scale indicates the value of $\frac{\ln{N_{[r_0]}} }{\ln(r_0)}$. In order to test the fractal behaviour of a structure, we can evaluate the compatibility between {the productivity of the structures in NGC$~$2264 data} and the productivity of a fractal structure with a similar fractality. 

This can be done easily as the slope of the productivity depends directly on our model parameters. Moreover, in the case of an ideal fractal cascade, the measurement of the fractality corresponds to the fractal index we need to put in our model. Therefore, measuring the average fractality per graph-source in a structure gives us what should be the fractal index, which we put in the model for comparison.  

\subsubsection{Fractality uncertainty and limit of the model}
\label{Subsubsection:ModelLimit}

In this subsection, we aim to characterise the uncertainty of the fractility coefficients computed from our model, which we want to compare to the theoretical fractality index of an ideal cascade process. Unlike the former network, the ideal cascade is characterised by a fractional number of substructures for a given scale. The network built from the NGC$~$2264 data or our model is characterised by an integer number of substructures for a given scale. Therefore, for each individual hierarchical structure, we derive an approximate value of the fractality. However, if we assume {that a population of structures subdivide with the same fractal law} (e.g. generated by our model), {then taking the average of the fractality coefficients of this population must converge to the true value of the fractality.
}

\begin{figure}[!h]
    \centering
    \includegraphics[width=8cm]{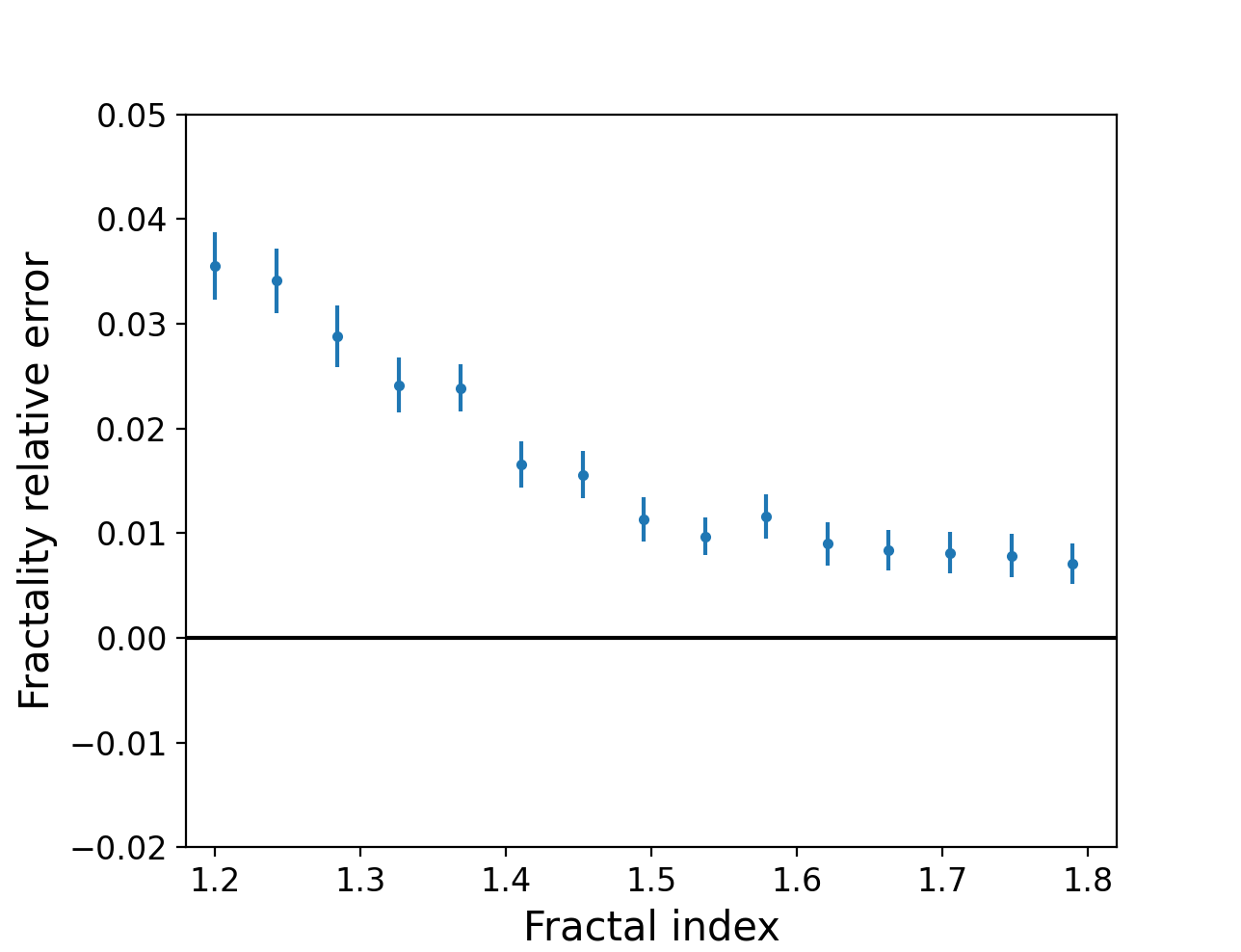}
    \caption{Average fractality relative error over the structures for different fractal index put in the model. For each fractal index we sample 1\,000 initial graph-sources, subdividing through levels defined by the scaling ratios of NGC$~$2264. Error bars are taken as $\sigma/\sqrt{N-1}$ where $\sigma$ is the  standard deviation of the fractality of the $N$ structures.}
    \label{fig:ferror}
\end{figure}

There is another source of error when considering model-generated data. Indeed, our model requires that a substructure at a level $l$ subfragments into at least one child for the cascade to continue towards lower levels. This condition guarantees that the resulting network is continuous along the branches, and that each branch is equivalent from one to another. For example, consider structures that are supposed to contain 1.2 substructures at the level $l = 1$ and 1.5 substructures at the level $l = 2$ on average, and specifically any graph-source of one of the structures. At the level $l,$ the model selects an integer number of substructures to put into the level $l = 1$: either 1 or 2 can be selected, but on average it has to be 1.2. Assuming the model selects 2, we need to place 1.5 substructures at the level $l = 2$ inside two existing substructures at level $l = 1$. As zero children is impossible, the model always puts one substructure inside each of the existing substructures. The choice of selecting two substructures at the level $l = 1$ always results in two substructures at the level $l = 2$. However, the choice of selecting one substructure at the level $l =1$  results in 1.5 substructures on average at the level $l = 2$. The total number of substructures tends to be overestimated. As a consequence, the fractality is overestimated. This can be seen in Figure \ref{fig:ferror} where the value of the fractality is systematically above the expected value for model-generated data. However, this error is always less than 5\% of the expected value. This effect is due to a low fractal index coupled to a low scaling ratio between two consecutive levels.

\begin{figure*}
   \centering
\begin{tabular}{l|cccccc}
Geometry&
\includegraphics[width=2cm]{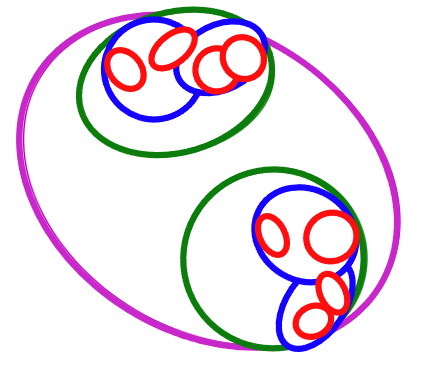}&
\includegraphics[width=2cm]{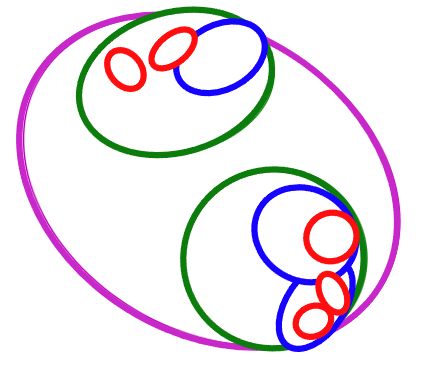}&
\includegraphics[width=2cm]{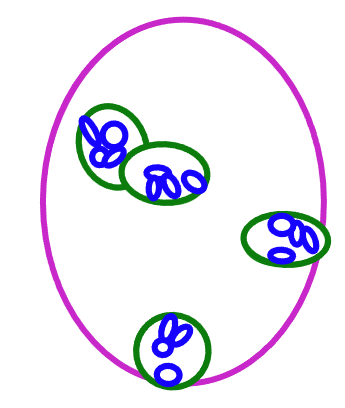}&
\includegraphics[width=2cm]{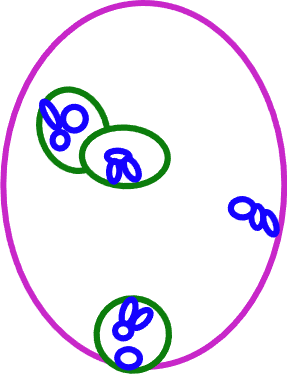}&
\includegraphics[width=2cm]{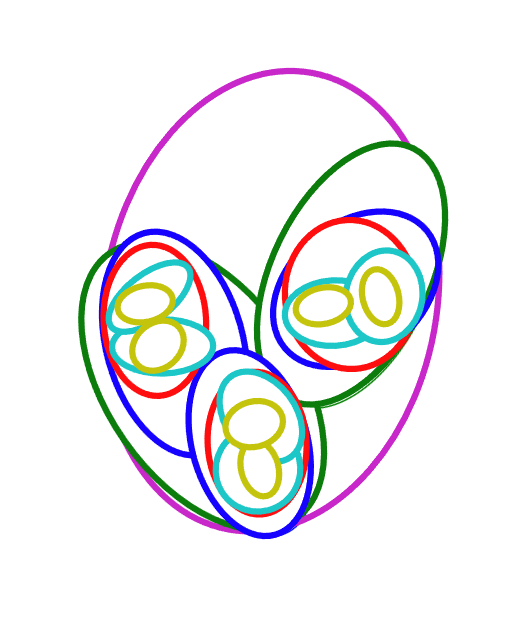}&
\includegraphics[width=2cm]{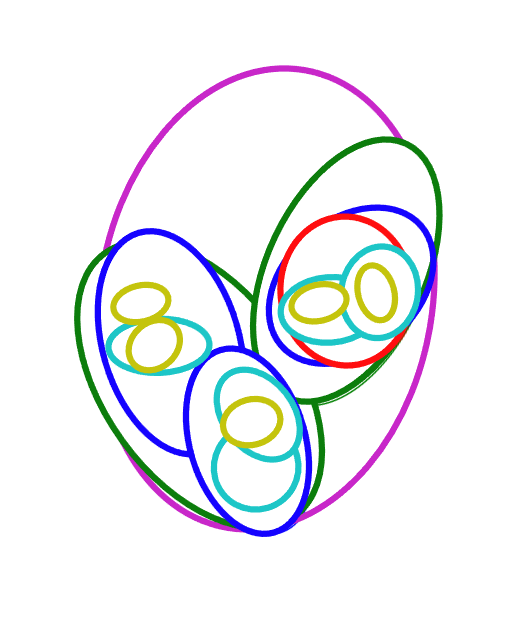}\\
Network&
\includegraphics[width=2cm]{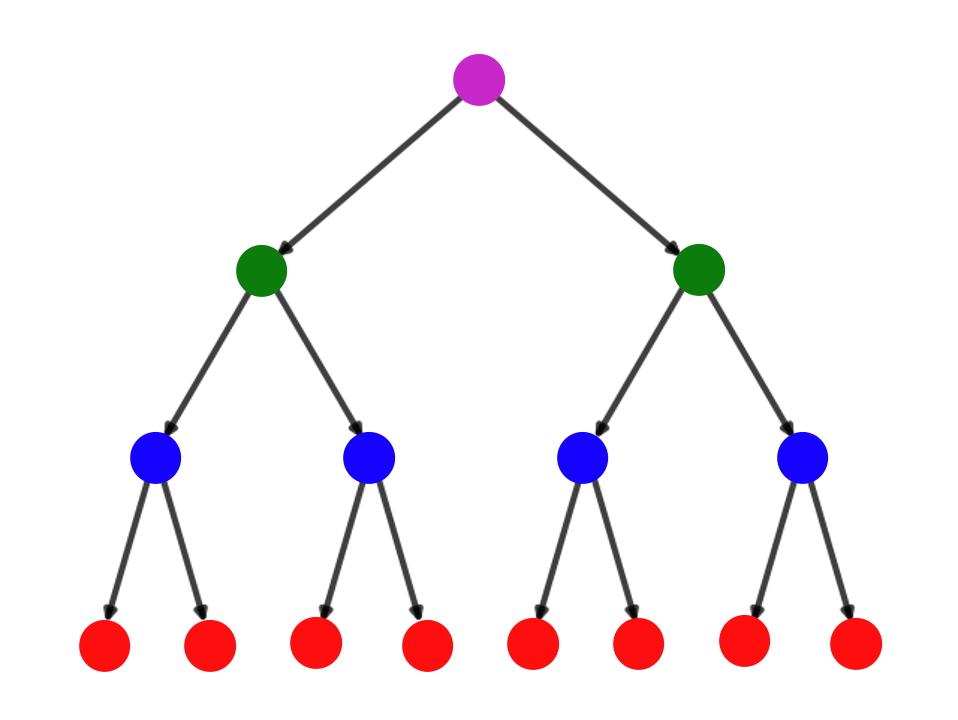}&
\includegraphics[width=2cm]{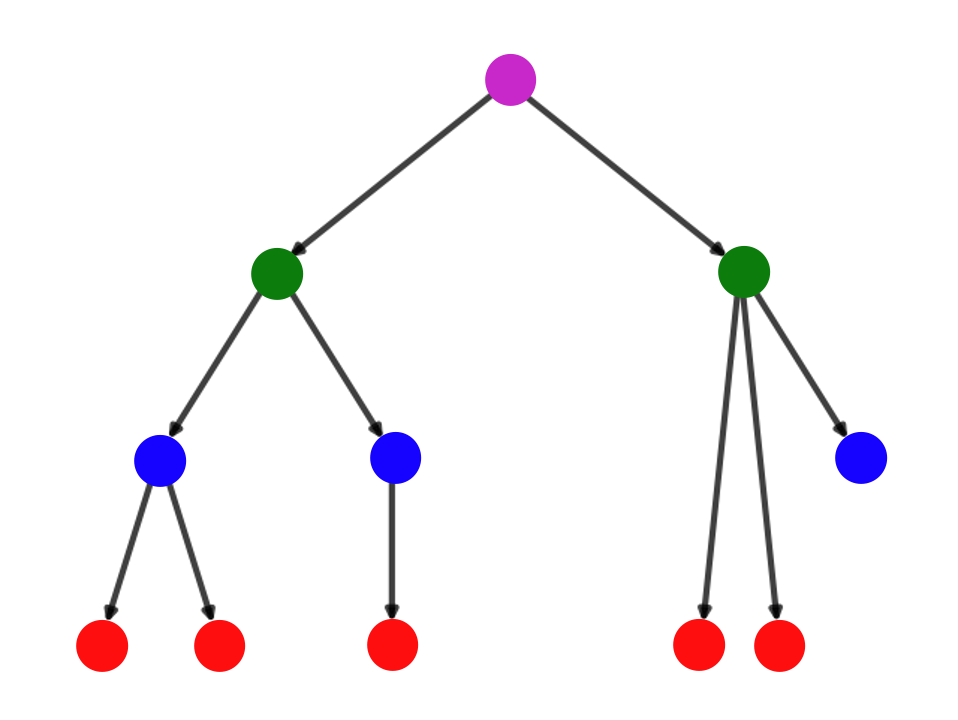}&
\includegraphics[width=2cm]{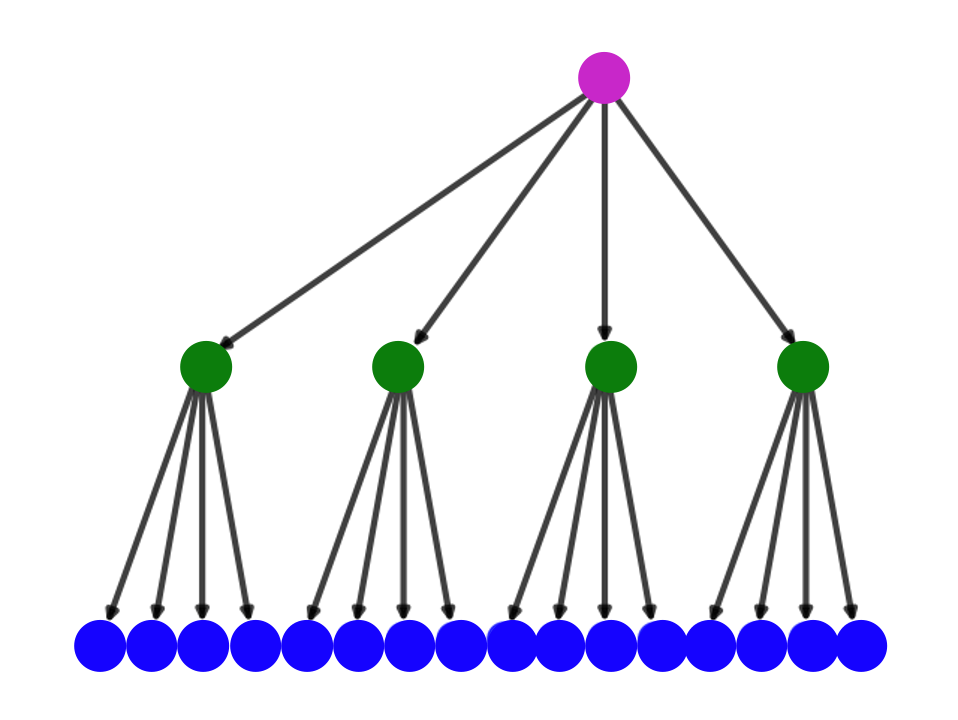}&
\includegraphics[width=2cm]{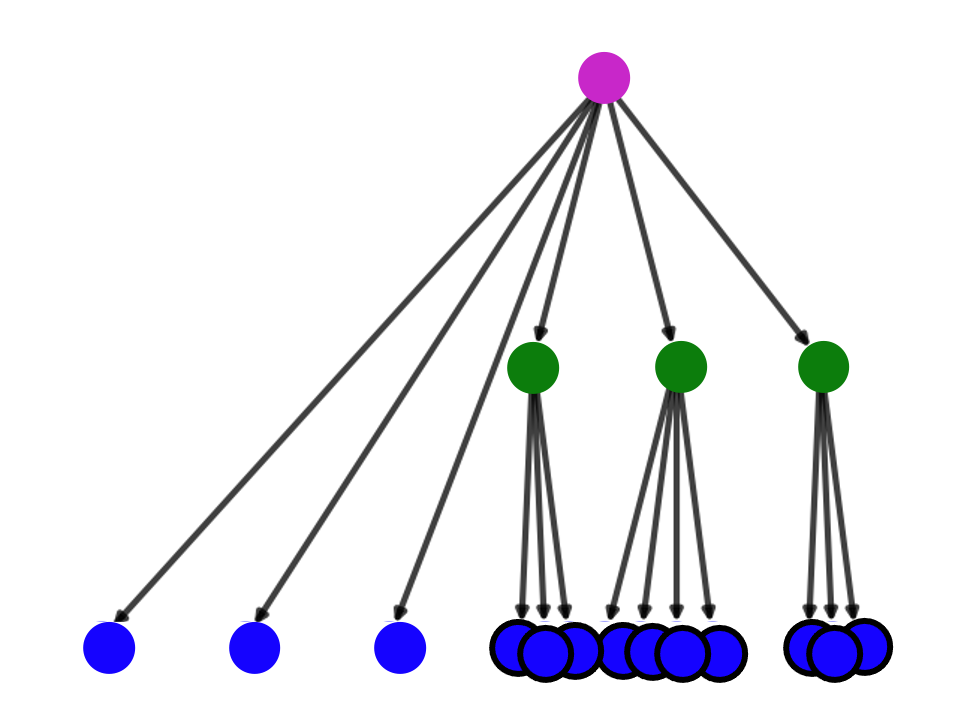}&
\includegraphics[width=2cm]{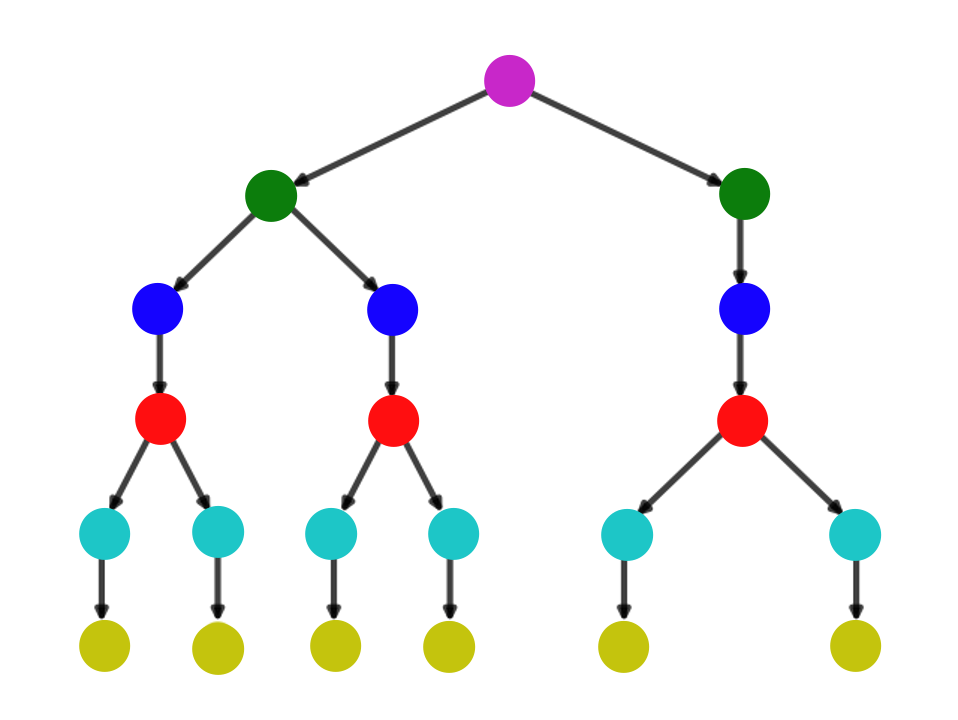}&
\includegraphics[width=2cm]{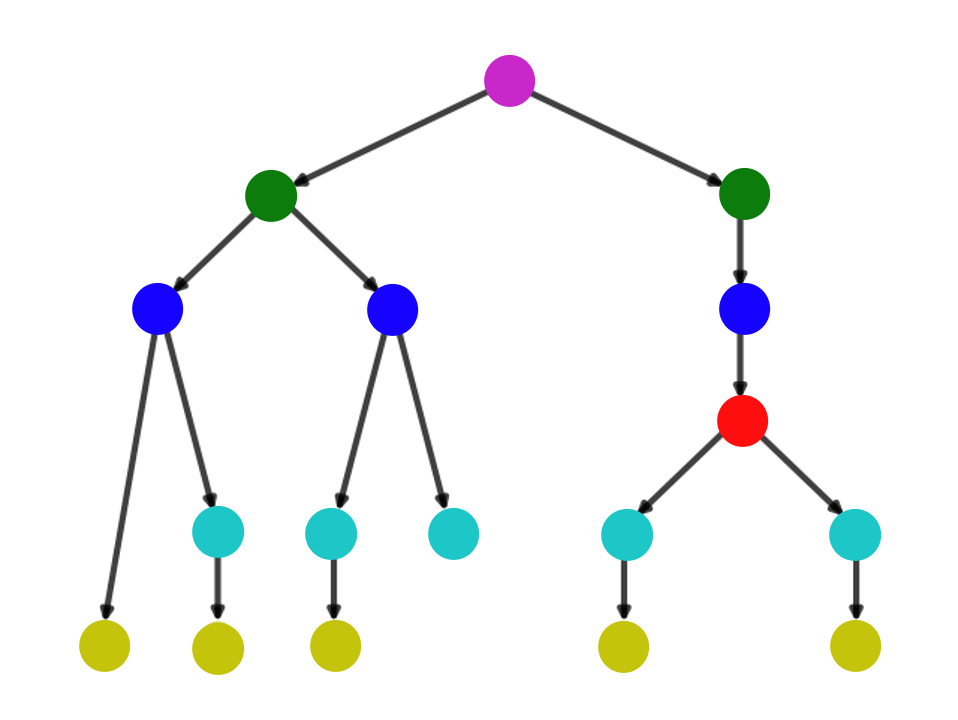}\\
\hline
$r_{l \rightarrow l+1}$&
2   &   2   &   4   &   4   &   1.5 &   1.5 \\
node removed&
0   &   4   &   0  &   4    &   0   &   4 \\
\hline
\hline
$\mathcal{F}$ computed&
2.0   &   1.7   &   2.0   &   1.9   &   2.0   &  1.8 \\
\end{tabular}

    \caption{Example of structures generated by the model with a scaling index $N_{[2]} = 2$ and multiple constant scaling ratio $r_{l\rightarrow l+1} = r, \forall l$. Nodes were removed randomly to simulate the types of structures that the observation can provide. The fractality coefficient $\mathcal{F}$ is reported as well as the number of missing nodes. In all of these generations, $\mathcal{F} = 2$ is expected.}
    \label{table:model}
\end{figure*}

\begin{figure}[!h]
    \centering
    \includegraphics[width=9cm]{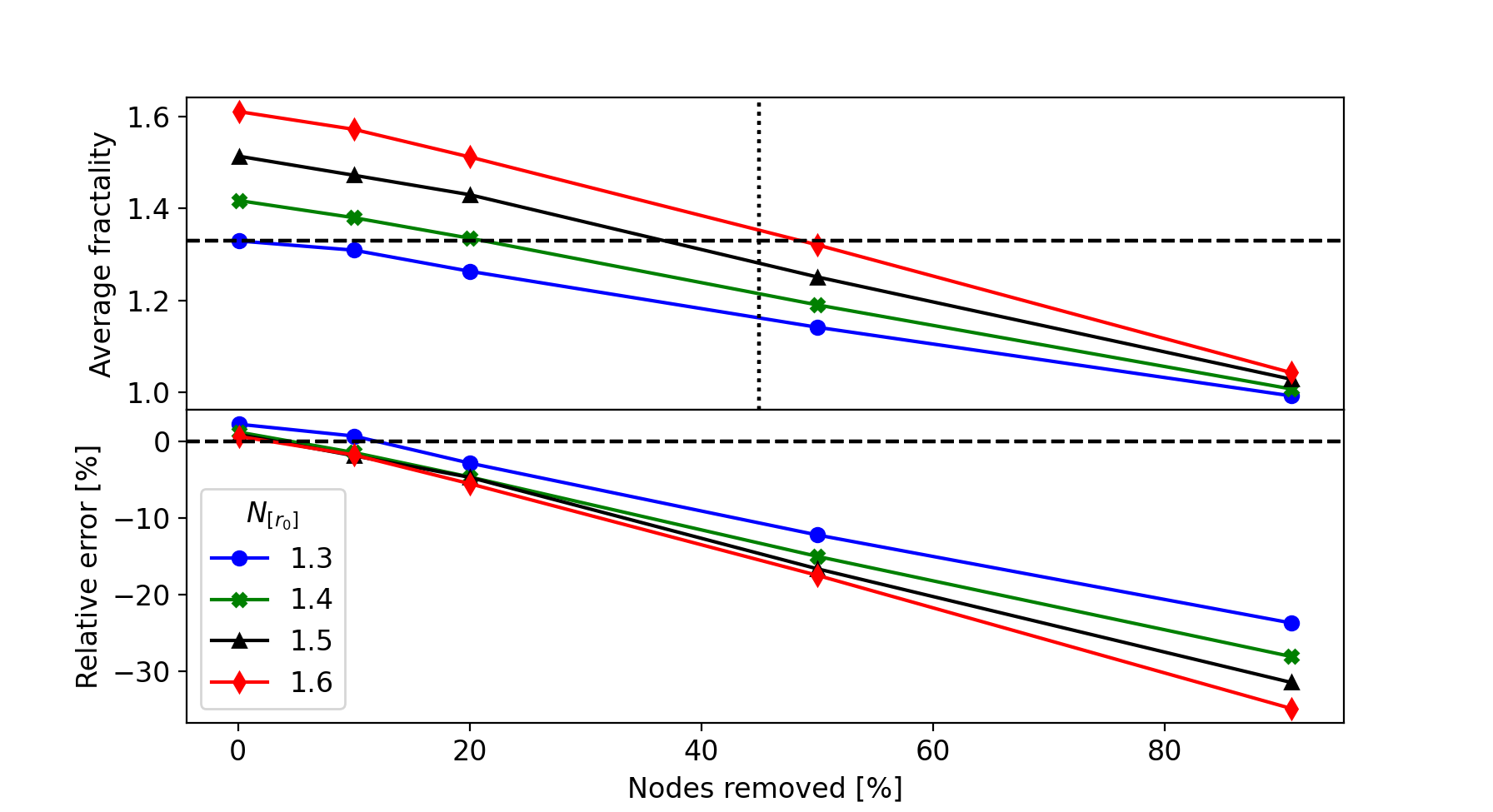}
    \caption{Ten thousand graph-sources sampled in our model and subdivided along the scales defined for NGC$~$2264 (see Section \ref{Subsection:Data}) for the four different fractal indices $N_{[r_0]}$ reported in the legend. \textit{Top}: Average fractality of the whole population for each fractal index as a function of the removed nodes. Horizontal dashed line indicates the fractality measured for NGC$~$2264 in Section \ref{Subsection:FractalHierarchiInNGC}. The vertical dashed line corresponds to the estimation of the maximum proportion of nodes undetected in NGC$~$2264 in Section \ref{Subsection:FractalHierarchiInNGC}. \textit{Bottom}: Relative error of the fractality as a function of the removed nodes. Horizontal dashed line indicates a 0\% variation. Positive relative errors are due to over-sampled substructures in the model with low fractal index (more details in text Section \ref{Subsubsection:ModelLimit}).}
    \label{fig:uncert}
\end{figure}

If we want to compare our model with the data for NGC$~$2264, we need to take into account another effect. Indeed, in these data, some of the intermediate substructures might not be detected at a specific scale. An undetected clump at this scale directly affects the fractality value because it reduces the number of nodes of  the
structure by adding a hole in the cascade (see Figure \ref{table:model}). Our model can be used to quantify the error caused by the missed detections to determine a fractality confidence interval. For different fractal index (see Figure \ref{fig:uncert}), we use the scaling ratios that are defined by the NGC$~$2264 data set and we generate a population of 1\,000 graph-sources. We then randomly delete a specific proportion of nodes inside each structure, excluding their graph-sources, and compute the fractality.

\subsection{Fractal hierarchy in NGC$~$2264}
\label{Subsection:FractalHierarchiInNGC}

Further analysis regarding the hierarchical structures is performed with NGC$~$2264 data using both our geometrical model and the metric $\mathcal{F}$ defined in Section \ref{Subsection:Metrics}. This analysis is performed specifically on hierarchical structures because both the model and the metric we designed were made to describe hierarchy. Computing the fractality on linear structures systematically returns a value of  1, and computing it for isolated {structures} is useless by definition. As pointed out in Section \ref{Subsubsection:ModelLimit}, fractality computation is biased by the fact that some hierarchical structures may miss intermediates substructures. In order to properly compare the structures of NGC$~$2264  with the model, we need to estimate the number of missing substructures in these structures. As the networks that represent these structures contain holes at some levels instead of nodes, we count the number of holes. Hence, a structure containing $\{1,0,0,2\}$ substructures at a level $l$ is considered to possess four holes in total (two holes per graph-sink). Under the assumption that no graph-sink is missing, and in the worst case scenario, this structure can have a $\{1,2,2,2\}$ configuration. This method overestimates the number of holes, because it does not consider the eventual merging of substructures at higher scales. As we overestimate the number of substructures, we overestimate the fractality we want to compute. Hence, we can assess an upper limit. To estimate a lower limit, we take the fractality value without any hole-filling. \\

\begin{figure}[!h]
\centering
\includegraphics[width=8cm]{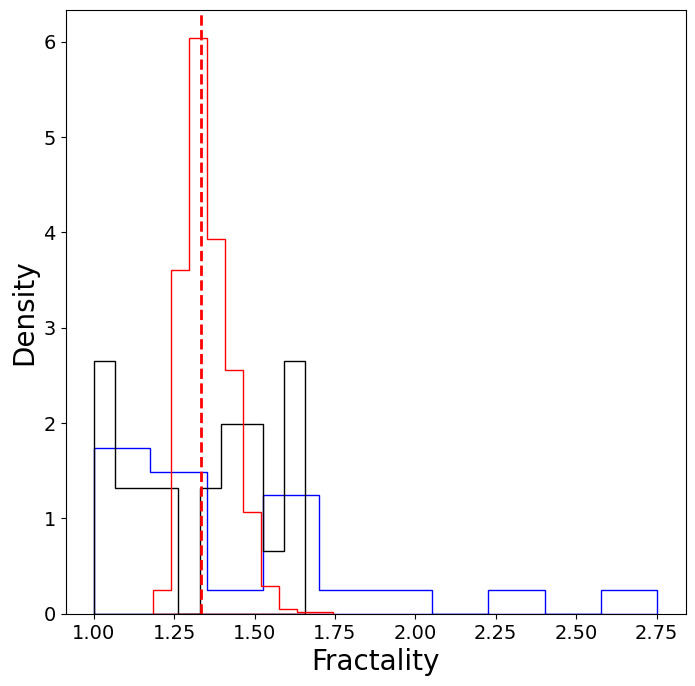}
\caption{{Normalised distribution of fractality coefficient $\mathcal{F}$ for different sets of data.} In blue, $\mathcal{F}$ distribution for the hierarchical structures in NGC$~$2264. In red, $\mathcal{F}$ distribution with our fractal model of index 1.57 where we have removed 45\% of the nodes. In black, $\mathcal{F}$ distribution for the convolved data (see Section \ref{Section:Discussion}). The red dotted line corresponds to the median value of 1.33.}
\label{Figure:distribproductivity}
\end{figure}

On average, the extracted hierarchical population for NGC$~$2264 gives $\mathcal{F} = 1.45$. However, the median value (1.33) seems to be a more appropriate estimator because of the asymmetry of the distribution (see Figure \ref{Figure:distribproductivity}). As our data contain holes, this estimation of the fractality is underestimated. We therefore take $\mathcal{F}_{min} = 1.33$ as a lower limit for the fractality. We estimate the maximum proportion of holes in hierarchical structures to be  45\%. This corresponds to a fractality negative variation of $\sim$15\% according to Figure \ref{fig:uncert}. Solving $(1-0.15) \mathcal{F}_{max} = 1.33$, we estimate the upper limit $\mathcal{F}_{max} = 1.57$. This value is also consistent with the median of the fractality distribution after filling the holes in the structures. As the true fractality for the whole population is likely to fluctuate between 1.33 and 1.57, we derive $\mathcal{F} = 1.45 \pm 0.12$ for the total population of hierarchical structures in NGC$~$2264. 

Taking into account the bias introduced by the holes, we choose to compare the NGC$~$2264 hierarchical structures with the structures of our fractal model of index 1.57, where we randomly deleted 45\% of the nodes inside each structure. In particular, the distribution of fractality coefficient in NGC$~$2264 is wider than the distribution extracted from our model (see Figure \ref{Figure:distribproductivity}). This means that the NGC$~$2264 structures and our model are not equivalent in terms of fractality despite our correction concerning the holes. In fact, a single value of fractality fails to describe every single structure in the data. If the fractal assumption were true, we would expect to get a homogeneous fractality coefficient throughout the cloud regardless of the scale considered (assumption of mono-fractal process) and also independent of the physical environment (temperature, density, turbulence). This dispersion may suggest that the fractality of {structures}, and by extension {the multiplicity of their substructures}, could depend either on local physical conditions or on the scale itself. \\

\begin{figure}[!h]
\centering
\includegraphics[width=8.5cm]{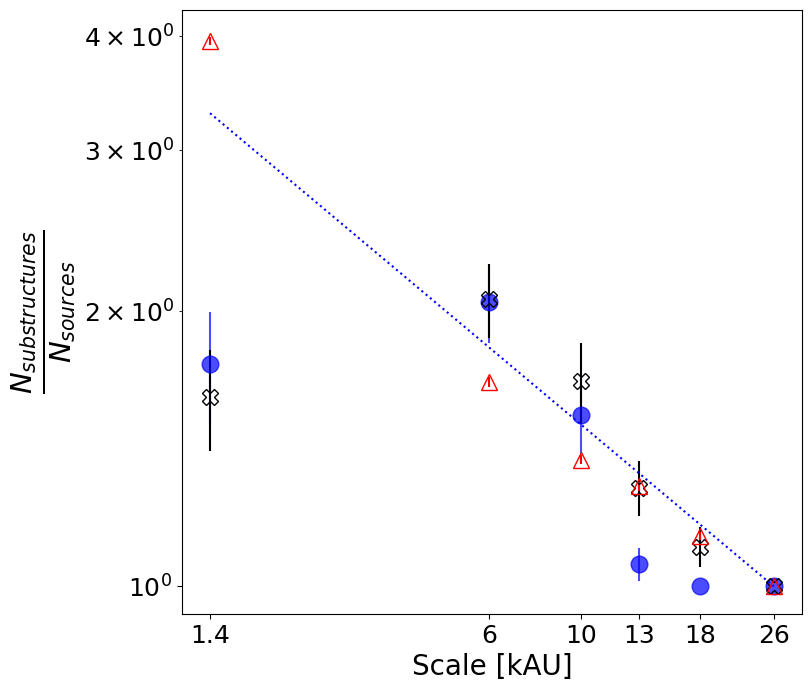}
\caption{Average productivity of hierarchical structures for each scale. Blue circles represent NGC$~$2264 data. Red triangles represent a population of 1\,000 structures generated by our geometrical model with a fractal index of 1.57 where we have removed 45\% of the nodes. Blue dotted curve represents an ideal fractal fragmentation of index 1.33. Black crosses represent the same curve for convolved data (see Section \ref{Section:Discussion}).}
\label{Figure:productivity}
\end{figure}

The hierarchical property of the structures of NGC$~$2264  can be tested by analysing the hierarchical cascade scale by scale using the productivity of {structures} per graph-source as introduced in Section \ref{Subsection:Productivity}. The result of the productivity curve in NGC$~$2264 is also compared with our model in Figure \ref{Figure:productivity} with a fractal index of 1.57 with 45\% of the nodes removed. This comparison confirms that the hierarchical structures in NGC$~$2264 are incompatible with a single fractal fragmentation. Indeed, between 26$~$kAU and 13$~$kAU, the productivity of the structures of NGC$~$2264  remains constant: the networks are locally linear until the resolution is high enough to observe a subdivision. No hierarchical cascade seems to occur at these scales. Because we do not obtain the same trend with our geometrical model, this effect is not due to low scaling ratios between the scales. The geometrical constraints coupled with random deletion of  holes cannot explain this result. It is important to consider the fact that we encounter the same issue as before concerning the potentially missed intermediate substructures. The previous technique that consists of filling the holes cannot be used here because it assumes that the hierarchy begins at the highest levels in order to assess what would be the worst-case scenario to evaluate an uncertainty, while here we investigate the multiplicity property scale by scale. A similar correction would completely bias the result, given that we do not know whether or not the hierarchical cascade actually has a starting scale; and if it does, we do not know what this starting scale would be. Therefore, we consider the curve without any correction and discuss its reliability in Section \ref{Subsection:CascadeInterpretation}. A depletion of YSOs can be seen  below 6$~$kAU: structures in NGC$~$2264 do not produce as many substructures as the model at the last level. Indeed, down to 1$~$kAU, the model predicts about four class 0/I YSOs per graph-source in a structure, whereas about one or two YSOs are produced per graph-source in most structures in NGC$~$2264. There are multiple factors that could explain this effect: {half of the graph-sinks are clumps which might indicate that the gas reservoir is not fully depleted, and star formation can still occur in the future ; the subdivision rate slows down to produce YSOs ; the oldest class 0/I evolve towards class II objects and are not taken into account in this study.}



\section{Discussion}
\label{Section:Discussion}

\subsection{Origin of the plateau observed in the hierarchical cascade}
\label{Subsection:CascadeInterpretation}

In Section \ref{Subsection:FractalHierarchiInNGC} we show that the hierarchical cascade in NGC$~$2264 is not fractal over the full 6-26$~$kAU range. Indeed, the substructure production scale by scale plateaued on scales of > 13$~$kAU. We perform a series of tests allowing us to evaluate the robustness of the cascade and check whether the plateau may arise from spatial resolution and/or wavelength effects. \\

\subsubsection{Spatial resolution effect}
\label{subsubsection:Spatialresolution}

First, we start by checking spatial resolution effects. When a cloud is observed, the flux received at a given wavelength is convolved with the instrumental beam whose size defines the angular resolution. Through convolution, two substructures at a given scale may or may not merge. If the substructures are spaced far enough apart such that they do not merge, we would expect to have a one-to-one correspondence between the original substructures and their convoluted counterparts, that is, a plateau in terms of substructure production. If the substructures are close enough to merge at higher resolution, then this plateau cannot exist as there is no longer a one-to-one correspondence. In order to evaluate the influence of spatial resolution on our results, we performed a convolution of the 160$~\mu$m emission image  at $[18.2, 24.9, 36.3]$" resolution. New catalogues of clumps were obtained using \texttt{getsf} source extraction on the new images using the same setup as for the original data (see Section \ref{Subsection:Data}). These catalogues were then reprocessed by our network procedure in order to extract the new structures, keeping the original 70$~\mu$m and class 0/I YSO data at lower scales. When we investigate the substructure production scale by scale with these modelled data, we do not observe any plateau on scales > 13$~$kAU, but we see an actual increase in the substructure production (see Figure \ref{Figure:productivity}). Therefore, the plateau cannot originate from a pure spatial resolution effect.

\subsubsection{Wavelength effect}

Secondly, we discuss the influence of the wavelength on the detection of clumps. Indeed, each spatial resolution is associated with a specific wavelength, and each wavelength is associated with a different flux and emission temperature of dust. We do not observe exactly the same sky between images at 500$~\mu$m and images at 250$~\mu$m or at 160$~\mu$m. The plateau that is observed for scales larger than 13$~$kAU (see Figure \ref{Figure:productivity}) could be due to the failure to detect large-scale clumps that would allow their productivity curve to flatten at high scales. We also investigated whether there is indeed a tendency to loose clumps at large scales. \\

Let us set I($\theta$, $\lambda$) as the map associated with the ${\theta}$ resolution in arcsec and ${\lambda}$ wavelength in$~\mu$m for simplicity. We already have original data for which a wavelength is associated with a resolution (i.e. I(36.3, 500), I(24.9, 350), I(18.2, 250), I(13.5, 160)), and the data we obtained by convolving the 160$~\mu$m map to higher resolutions (i.e I(13.5-to-36.3, 160)). In order to complete these two data sets, we additionally convolved the 250 and 350$~\mu$m maps at 36.3" resolution and extracted clumps with the \texttt{getsf} procedure. We thus obtained new I(36.3, 250) and I(36.3, 350) data. Using our methodology of associating clumps with each other, we can connect the I(36.3, 160-to-500) data in terms of wavelength (and not in terms of scales). With these data we investigated the persistence of the clumps in each of the data sets, which is the proportion of clumps that are found from one scale or wavelength to another. This study was performed in terms of constant wavelength (I(13.5-to-36.3, 160)), constant resolution (I(36.3, 160-to-500)), and real data (I(13.5-to-36.3, 160-to-500)). We completely ignored the 70$~\mu$m clumps and YSOs because we only want the large-scale information and, unlike longer wavelengths, 70$~\mu$m does not trace optically thin thermal emission. 
We want to assess how the appearance of a clump changes along the wavelengths. As we are convolving the images to have new datasets and populations of clumps, we want to avoid the case where a clump in the vicinity of the one we are looking at interferes in any way with its persistence measurement. As hierarchical structures are a complex of clumps, and isolated structures are observable in only one wavelength, we need to consider only the linear structures. Indeed, the latter represent multi-scale and multi-wavelength single objects without immediate proximity of another object. With this method of synthesising data at several different wavelengths (I(36.3, 160-to-500)), we implicitly couple the resolution effect with the wavelength effect. Indeed, with successive convolutions of a map, the flux of a clump is diluted in its close environment, which corresponds to the size of the Gaussian kernel with which we convolve the map. Despite this, we know what happens during a pure convolution with the set I(13.5-to-36.3, 160) (see Section \ref{subsubsection:Spatialresolution}). \\

\begin{figure}[!h]
    \centering
    \includegraphics[width=9cm]{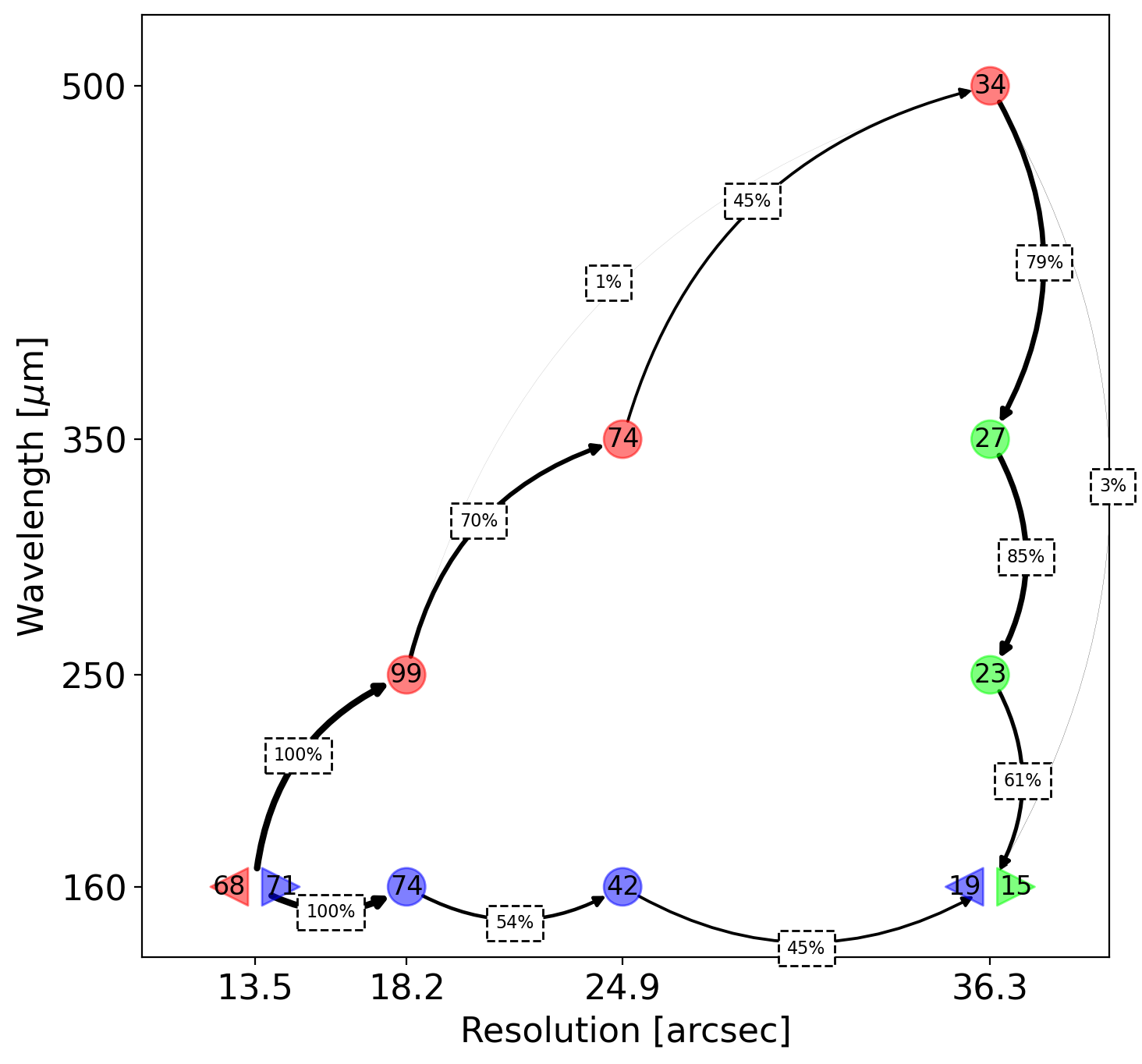}
    \caption{Representation of the successive association of clumps between scales and wavelengths for linear structures. Red markers are associated with the original data from NGC$~$2264. Blue markers are associated with the data from the convolution of the image at 160$~\mu$m to resolutions of 13.5, 24.9, and 36.3". Green markers are associated with data from the convolution of the 160, 250, and 350$~\mu$m maps to a resolution of 36.3". The numbers in the markers indicate the number of associated substructures. Triangle markers are turning points to make it clear that even though the data are the same, the substructures associated with the linear structures may change. The percentage in the middle of the arrows gives the proportion of substructures that persist from the out-coming node to the in-coming node.}
    \label{fig:persistence}
\end{figure}

Comparing the persistence of I(13.5-to-36.3, 160-to-500) and I(13.5-to-36.3, 160) data, we show (see Figure \ref{fig:persistence}) that {the persistence of the original data is very similar to the persistence of data obtained from the convolution}. For the 24.9" to 36.3" transition in both data sets, 45\% of the clumps at 24.9" can be retrieved at 36.3". In the other transition, 70\% of the I(18.2, 250) clumps are retrieved in the I(24.9, 350) data whereas 54\% of the I(18.2, 160) clumps are retrieved in the I(24.9, 160) data. This suggests that from one scale to another, on average, the resolution effect dominates the probability of detecting a clump, and wavelength might influence the detection of clumps. Nevertheless, we systematically detect more clumps at equal resolution in I(13.5-to-36.3, 160-to-500) than in I(13.5-to-36.3, 160). Therefore, if there is a wavelength effect, it tends to increase the number of clumps detected. However, in order to flatten the productivity curve at large scales, the number of substructures needs to be reduced so that a plateau can be obtained. Furthermore, the reason we have a plateau in the productivity curve is that a single substructure is produced between 250 and 500$~\mu$m. There should be a loss of detection in this area. If we look at the persistence on the I(36.3, 160-to-500) data, by degrading the map at 350$~\mu$m to 36.3" resolution, we find 79\% of the clumps detected at 500$~\mu$m. Therefore, if we have a 21\% loss between 500 and 350$~\mu$m, this means that one clump out of five disappears. To explain the plateau, we would need one clump out of two given the geometrical constraints, which is at least a 50\% loss. As this loss is not obtained at longer wavelengths, the lack of clumps on this range seems unlikely. The same applies to losses between 350 and 250$~\mu$m. On the other hand, if there were a tendency for clumps to appear at longer wavelengths, this could not explain a flattening of the curve because there would be more substructures in this wavelength domain. Wavelength effects appear to be negligible compared to resolution effects and seem insufficient to explain a flattening of the hierarchical cascade at large scale. The spatial resolution effects are already contained in the 160$~\mu$m convolved data set, and are already contained in the productivity curve in Figure \ref{Figure:productivity}. We can therefore assess the physical reality of the plateau observed in the productivity curve for scales of between 13$~$kAU and 26$~$kAU.

\subsection{Interpretation of the hierarchical cascade}

The hierarchical cascade we outlined can be interpreted as a fragmentation cascade. According to this framework, the turbulent \citep{larson_calculations_1978, padoan_supersonic_1995} or \textit{gravo-turbulent} \citep{hopkins_general_2013, guszejnov_universal_2018, vazquez-semadeni_global_2019} fragmentation model assumes that the gravitational instabilities responsible for the cloud fragmentation originate from supersonic \citep{padoan_supersonic_1995} velocity propagation. These turbulent motions allow matter to condense locally, resulting in a cascade of turbulence from large to small scales. These instabilities subsequently `subfragment', allowing a chain of `collapse within collapse' \citep{vazquez-semadeni_global_2019, krause_physics_2020} in which the smaller scales accrete matter from the larger scales. However, this idea of a hierarchy in the fragmentation process is not universal. Indeed, as the gas flow velocities become subsonic (Mach number > 2-3, \citealp{hopkins_general_2013, guszejnov_isothermal_2018}), the cloud may collapse into a single massive fragment instead of forming several sibling fragments \citep{guszejnov_isothermal_2018}. Therefore, if we consider that the hierarchical cascade originates from fragmentation processes, we can dissociate two regimes. Indeed, in this interpretation, hierarchical structures describe a mode of hierarchical fragmentation (e.g. \citealp{hoyle_fragmentation_1953, vazquez-semadeni_global_2019}) in which a single clump fragments into multiple pieces while linear structures may be associated to a monolithic fragmentation where the collapse is radially concentrated and one clump collapses into a single clump or YSO (e.g. \citealp{shu_self-similar_1977, krause_physics_2020}). 

This concept of hierarchical `subfragmentation' originates from \cite{hoyle_fragmentation_1953} who studied the hierarchical collapse from Galactic scale down to prestellar cores assuming isothermal collapse of a spherical cloud. In this model, we expect a fractal process of fragmentation in which two fragments are formed each time the scale of the parent is reduced by a factor of two. Applied to our work, this process would be associated to a fractality value of two. Considering a fragmentation process that starts at 13kAU, a fractality of two would, on average at 6kAU,  induce 2.17 fragments per graph-source whereas we actually get 2.05 $\pm$ 0.20 fragments at 6kAU in NGC$~$2264 which is consistent with the hierarchical model of \cite{hoyle_fragmentation_1953}. Nevertheless, hierarchical fragmentation is associated with turbulent scale-free processes which one would expect to exhibit fractal behaviour. However, there does not seem to be any fractal subdivision in NGC$~$2264, but rather a range of scales for which the cloud does not fragment before starting these hierarchical processes. Therefore, it appears that turbulent processes alone do not fully describe the hierarchical cascade observed in NGC$~$2264. \cite{hopkins_general_2013} modelled intermittency phenomena that can originate with shocks or sound waves. These intermittency phenomena have the effect of constraining the hierarchical cascade process between several scales, which could explain the plateau for scales  > 13kAU seen in the hierarchical cascade in NGC$~$2264. Consequently, it would be necessary to develop a multi-fractal description of the cloud, where the hierarchical cascade is not scale-free but scale-dependent.

\subsection{Influence of star formation history}

Regarding the stellar component of the hierarchical structures, it appears that the production of YSOs slows down below a certain scale of < 6kAU. However, it is also possible that this is an incompleteness effect with respect to the star formation phenomena that occurred in the past in NGC$~$2264. Indeed, we have only taken into account the class 0/I YSOs, leaving aside the more evolved class II YSOs. Knowing that the hierarchical structures are mainly located in the central hubs, it is likely that the available gas reservoir replenishes throughout the filaments (\textit{conveyor belt} model, \citealp{longmore_formation_2014}) which would allow the production of YSOs for several star forming events. This hypothesis would favour the cohabitation of class 0/I with class II YSOs within the hierarchical structures. However, this scenario does not seem to be able to fully explain this decrease in star production because, taking into account the class II YSOs from \cite{rapson_spitzer_2014}, the hierarchical structures would produce on average 2.26 $\pm$ 0.28 YSOs per graph-source (instead of 1.75 $\pm$ 0.24 with class 0/I only). Considering the \cite{hoyle_fragmentation_1953} fragmentation estimation (fractality $\mathcal{F}$ = 2), we expect to produce on average nine YSOs per graph-source. Therefore, the fragmentation of the clumps into YSOs seems to be inhibited, or is still in the process of forming prestellar objects. 

\section{Conclusion}
\label{Section:Conclusion}

We designed a methodology based on connected network representation to link clumps of different scales and YSOs. This linkage respects an inclusion principle in order to study a snapshot of the spatial properties of the hierarchical cascade. The computed network is structured in different levels, each one of them being associated with a specific spatial scale. We designed a geometric model to generate a population of extended objects that subdivide according to a fractal law whose associated fractal index takes into account the scaling ratios between the levels. Using the Herschel five-band observation of resolution [8.4, 13.5, 18.2, 24.9, 36.3]" coupled with the class 0/I YSOs from the Spitzer survey of NGC$~$2264, we applied our methodology and extracted structures of three types: hierarchical, which we associate with a hierarchical fragmentation process; linear, which we interpret as a monolithic fragmentation process; and a final isolated mode. Our analysis of the structures reveals various phenomena. {We observed that linear structures that incubate a YSO exist, showing that monolithic collapse and hierarchical collapse can coexist within the same cloud. In particular, the hierarchical structures are the main incubators of YSOs in NGC$~$2264, and have therefore been the main fuel for star formation. Finally, regions with high column density are correlated with hierarchical collapse while other regions are dominated by monolithic collapse.}


Our network analysis and comparison with our model show that {although we were able to measure an average fractality of $\mathcal{F}$ = $1.45 \pm 0.12$ on the whole network, the hierarchical cascade is incompatible with a purely fractal model when we analyse it scale after scale. We also find that no hierarchical cascade occurs between 26 and 13$~$kAU, and that no obvious hierarchical cascade seems to occur between 6$~$kAU and the YSO scale of class 0/I at 1.4kAU, indicating that hierarchical cascade may occur mainly between 6 and 13$~$kAU in NGC$~$2264.}

Our methodology provides a new approach to studying fragmentation processes within molecular clouds. We are planning future work to define a metric to quantify how the subdivision of our structures deviates from a scale-free, or fractal, process. However, the fractality coefficient we define can be reapplied to analyse the subdivision of dendrograms \citep{rosolowsky_structural_2008} in order to study the hierarchy in column density instead of hierarchy in scales as is done in the present paper. By studying the hierarchy in density using dendrograms, it is possible to investigate the cascade from a larger dynamic. 

One of the goals of this work is to provide tools to study the links between multi-scale structure of gas and the multiplicity of YSOs/NESTs. However, the current study is limited by the scale dynamics of the gas to which we have access, from 6$~$kAU to 26$~$kAU. In order to better investigate the link between multiple systems (< 1$~$kAU), the UWPs of high-order multiplicity (< 10$~$kAU) \citep{joncour_multiplicity_2017}, and the small stellar groups (NESTs, \citealt{joncour_multiplicity_2018}) together with the multi-scale hierarchical cascade of the gas, we would need to investigate scales smaller than 6$~$kAU. Indeed, at this stage we cannot explain the high multiplicity observed in these stellar structures which contain more than four YSOs each. In order to probe scales smaller than 6$~$kAU in NGC$~$2264, more observations need to be carried out using the NOrthern Extended Millimeter Array (NOEMA, the updated Plateau de Bure Interferometer \citealt{guilloteaunoema1992}), or the Atacama Large Millimeter Array (ALMA, \citealt{wooten_thompson2009}) facilities. \\

The application of clustering methods to our structures could also help to elucidate the close relationship between the spatial properties (separation, first neighbours, etc.) of stellar systems and their environment, which can be single clumps or clusters of clumps. Do small stellar groups arise from a simple hierarchical collapse of a clump, or are their spatial extent and multiplicity more compatible with an association of clumps that would result from the fragmentation of a filament, with each of these clumps in turn collapsing hierarchically or monolithically? Most probably, both scenarios exist \citep{offner2022}: the UWPs and large NESTs may result from the fragmentation of filaments (scale > 13$~$kAU), and the hierarchical fragmentation of clumps may proceed at smaller range, whereas disc fragmentation around YSOs may explain the very close stellar systems (scale < 100$~$AU). Our methodology is suitable for studying the spatial characteristics of hierarchical fragmentation, but gives no indication of the temporal aspects. However, the same metrics can be used to describe the temporal evolution of fragmentation in simulations.

\begin{acknowledgements} 
The authors would like to deeply and sincerely thank A. Men'shchikov for his mentorship regarding the learning process of handling \texttt{getsf} software. We are also grateful to the ECOGAL project (European Research Council synergy, Grant: 855130) and its members for all the discussion and workshop opportunities we benefit from. 
\end{acknowledgements}

\bibliographystyle{aa} 
\bibliography{biblio}

\end{document}